\documentclass[12pt]{article}

\usepackage{graphicx,amsmath,natbib,multicol}
\usepackage{amsfonts,amssymb,latexsym,textcomp,txfonts}
\usepackage{epsfig}
\usepackage{fullpage}
\usepackage{subfigure}
\usepackage{url}
\usepackage{times}

\newcommand{\bfbeta} {\boldsymbol{\beta}}

\newcommand{\bfmu} {\boldsymbol{\mu}}

\newcommand{\bflambda} {\boldsymbol{\lambda}}

\newcommand{\bfLambda} {\boldsymbol{\Lambda}}

\newcommand{\bfOmega} {\boldsymbol{\Omega}}

\newcommand{\bfZ} {\mathbf{Z}}
\newcommand{\bfY} {\mathbf{Y}}
\newcommand{\bfX} {\mathbf{X}}

\newcommand{\bfE} {\mathbf{E}}
\newcommand{\bfW} {\mathbf{W}}
\newcommand{\bfD} {\mathbf{D}}
\newcommand{\bfA} {\mathbf{A}}
\newcommand{\bfB} {\mathbf{B}}
\newcommand{\bfK} {\mathbf{K}}
\newcommand{\bfQ} {\mathbf{Q}}
\newcommand{\bfU} {\mathbf{U}}
\newcommand{\bfPsi} {\mathbf{\Psi}}
\newcommand{\bfPhi} {\mathbf{\Phi}}

\newcommand{\bfx} {\mathbf{x}}

\newcommand{\bfb} {\mathbf{b}}

\DeclareMathOperator{\diag}{diag}

\newcommand{\ground}{\mathcal{G}}

\newcommand{\nbd}{\mathsf{nbd}}
\newcommand{\normal}{\mathsf{N}}

\newcommand{\Poi}{\mathsf{Poi}}

\newcommand{\Uni}{\mathsf{Uni}}
\newcommand{\GWis}{\mathsf{Wis}}

\newcommand{\cone}{\mathsf{P}}

\title{Bayesian inference for general Gaussian graphical models with application to multivariate lattice data}
\author{Adrian Dobra\footnote{Departments of Statistics, Biobehavioral Nursing and Health Systems and the Center for Statistics and the Social Sciences, Box 354322, University of Washington, Seattle, WA 98195, U.S.A. (email: adobra@uw.edu).}, Alex Lenkoski\footnote{Institut f\"{u}r Angewandte Mathematik, Universit\"{a}t Heidelberg, Im Neuenheimer Feld 294, 69115 Heidelberg, Germany (email: alex.lenkoski@uni-heidelberg.de).}  and Abel Rodriguez\footnote{Department of Applied Mathematics and Statistics, University of California, Mailstop SOE2, Santa Cruz, California 95064, U.S.A. (email: abel@soe.ucsc.edu).}}
\date{}
\begin{document}
\maketitle
\begin{abstract}
 We introduce efficient Markov chain Monte Carlo methods for inference and model determination in multivariate and matrix-variate Gaussian graphical models. Our framework is based on the G-Wishart prior for the precision matrix associated with graphs that can be decomposable or non-decomposable. We extend our sampling algorithms to a novel class of conditionally autoregressive models for sparse estimation in multivariate lattice data, with a special emphasis on the analysis of spatial data. These models embed a great deal of flexibility in estimating both the correlation structure across outcomes and the spatial correlation structure, thereby allowing for adaptive smoothing and spatial autocorrelation parameters.  Our methods are illustrated using simulated and real-world examples, including an application to cancer mortality surveillance.
\end{abstract}

\section{Introduction}


Graphical models \citep{lauritzen_1996}, which encode the conditional independence among variables using an undirected graph, have become a popular tool for sparse estimation in both the statistics and machine learning literatures.  Implementing model selection approaches in the context of graphical models typically allows a dramatic reduction in the number of parameters under consideration, preventing overfitting and improving predictive capability.  In particular, Bayesian approaches to inference in graphical models generate regularized estimators that incorporate uncertainty of the model structure. \\
 \indent The focus of this paper is Bayesian inference in Gaussian graphical models \citep{dempster_1972} based on the G-Wishart prior \citep{roverato_2002,atay-kayis_massam_2005,letac_massam_2007}. This class of distributions is extremely attractive since it represents the conjugate family for the precision matrix whose elements associated with edges in the underlying graph are constrained to be equal to zero.  In addition, the G-Wishart naturally accommodates decomposable and non-decomposable graphs in a coherent manner. \\
\indent Many recent papers have described various stochastic search methods in Gaussian graphical models (GGMs) with the G-Wishart prior based on marginal likelihoods which, in this case, are given by the ratio of the normalizing constants of the posterior and prior G-Wishart distributions \--- see \citet{atay-kayis_massam_2005,jones_et_2005,carvalho_scott_2009,lenkoski-dobra-2010} and the references therein. Although the computation of marginal likelihoods for decomposable graphs is straightforward, similar computations for non-decomposable graphs raise significant numerical challenges that are a consequence of having to approximate the normalizing constants of the corresponding G-Wishart posterior distributions. In that regard, \citet{lenkoski-dobra-2010} point out that the Monte Carlo method of \citet{atay-kayis_massam_2005} is fast and accurate when employed for the computations of the normalizing constant of G-Wishart priors, although it is slow to converge for computations that involve G-Wishart posteriors. This remark leads to the idea of devising a Bayesian model determination method that avoids the computation of posterior normalizing constants.\\
\indent The contributions of this paper are twofold.  First, we devise a new reversible jump sampler \citep{Gr95} for model determination and estimation in multivariate and matrix-variate GGMs that avoids the calculation of posterior marginal likelihoods.  The algorithm relies on a novel Metropolis-Hastings algorithm for sampling from the G-Wishart distribution associated with an arbitrary graph. Related methods are the direct sampling algorithm of \citet{carvalho-massam-west-2007} and the block Gibbs sampler algorithm of \citet{piccioni_2000} and \citet{asci_piccioni_2006}. However, known applications of the former method involve exclusively decomposable graphs, while the latter method relies on the identification of all the cliques of the graph, a problem that is extremely difficult in the case of graphs with many vertices or edges \citep{tomita-et-2006}. Our method does not require the determination of the cliques and works well for any graph as it is based on the Cholesky decompositions of the precision matrix developed in \citet{roverato_2000,roverato_2002,atay-kayis_massam_2005}.  Second, we devise a new flexible class of conditionally autoregressive models (CAR) \citep{Be74} for lattice data that rely on our novel sampling algorithms.  The link between conditionally autoregressive models and GGMs was originally pointed out by \cite{BeKa95}.  However, since typical neighborhood graphs are non-decomposable, fully exploiting this connection within a Bayesian framework requires that we are able to estimate Gaussian graphical based on general graphs.  Our main focus is on applications to multivariate lattice data, where our approach based on matrix-variate GGMs provides a natural approach to create sparse multivariate CAR models.

\indent The organization of the paper is as follows. Section \ref{sec:ggm} formally introduces GGMs and the G-Wishart distribution, along with a description of our novel sampling algorithm in the case of multivariate GGMs.  This algorithm  represents a generalization of the work of \citet{giudici-green-1999} and is applicable beyond decomposable GGMs. Section \ref{sec:matrixggm} discusses inference and model determination in matrix-variate GGMs. Unlike the related framework developed in \citet{wang_west_2009} (which requires computation of marginal likelihoods and involves exclusively decomposable graphs), our sampler operates on the joint distribution of the row and column precision matrices, the row and column conditional independence graphs and the auxiliary variable that needs to be introduced to solve the underlying non-identifiability problem associated with matrix-variate normal distributions.  Section \ref{sec:spatialglm} reviews conditional autoregressive priors for lattice data and discuss their connection to GGMs.  This section discusses both univariate and multivariate models for continuous and discrete data based on generalized linear models.  Section \ref{sec:illust} presents three illustrations of our methodology: a simulation study, a spatial linear regression for modeling statewide SAT scores, and a cancer mortality mapping model.  Finally, Section \ref{sec:conclusions} concludes the paper by discussing some of the limitations of the approach and future directions for our work.

\section{Gaussian graphical models and the G-Wishart distribution} \label{sec:ggm}

We let $\bfX=\bfX_{V_{p}}$, $V_{p}=\{1,2,\ldots,p\}$, be a random vector with a p-dimensional multivariate normal
distribution $\normal_p(\mathbf{0},\bfK^{-1})$. We consider a graph $G=(V_{p},E)$, where each vertex $i\in V$ corresponds with a random variable $X_i$ and $E\subset V\times V$ are undirected edges. Here ``undirected'' means that $(i,j)\in E$ if and only if $(j,i)\in E$. We denote by $\mathcal{G}_{p}$ the set of all $p(p-1)/2$ undirected graphs with $p$ vertices. A Gaussian graphical model with conditional independence graph $G$ is constructed by constraining to zero the off-diagonal elements of $\bfK$ that do not correspond with edges in $G$ \citep{dempster_1972}. If $(i,j)\notin E$, $X_{i}$ and $X_{j}$ are conditional independent given the remaining variables. The precision matrix $\bfK=\left( K_{ij}\right)_{1\le i,j\le p}$ is constrained to the cone $\cone_G$ of symmetric positive definite matrices with off-diagonal entries $K_{ij}=0$ for all $(i,j)\notin E$.\\
\indent We consider the $G$-Wishart distribution $\GWis_G(\delta,\bfD)$ with density 
\begin{eqnarray} \label{eq:wishart}
p\left(\bfK|G,\delta,\bfD\right) & = &\frac{1}{I_G(\delta,\bfD)}(\mbox{det}\; \bfK)^{(\delta-2)/2}\exp\left\{-\frac{1}{2}\langle \bfK,\bfD\rangle\right\},
\end{eqnarray}
with respect to the Lebesgue measure on $\cone_G$
\citep{roverato_2002,atay-kayis_massam_2005,letac_massam_2007}. Here $\langle
\bfA,\bfB\rangle = \mbox{tr}(\bfA^{T}\bfB)$ denotes the trace inner product. The
normalizing constant $I_G(\delta,\bfD)$ is finite provided
$\delta>2$ and $\bfD$ positive definite \citep{diaconis_ylvisaker_1979}.\\
\indent We write $\bfK\in \cone_{G}$ as
\begin{eqnarray}\label{eq:decompk}
 \bfK = \bfQ^{T}\left( \bfPsi^{T}\bfPsi\right) \bfQ,
\end{eqnarray}
\noindent where $\bfQ=\left( Q_{ij}\right)_{1\le i\le j\le p}$ and $\bfPsi=\left( \Psi_{ij}\right)_{1\le i\le j\le p}$ are upper triangular, while $\bfD^{-1}=\bfQ^{T}\bfQ$ is the Cholesky decomposition of $\bfD^{-1}$. We see that $\bfK =\bfPhi^{T}\bfPhi$ with $\bfPhi = \bfPsi\bfQ$ is the Cholesky decomposition of $\bfK$. The zero constraints on the off-diagonal elements of $\bfK$ associated with $G$ induce well-defined sets of free elements $\bfPhi^{\nu(G)}=\{\Phi_{ij}:(i,j)\in \nu(G)\}$ and $\bfPsi^{\nu(G)}=\{\Psi_{ij}:(i,j)\in \nu(G)\}$ of the matrices $\bfPhi$ and $\bfPsi$ \citep{roverato_2002,atay-kayis_massam_2005}, where
$$
 \nu(G) = \{ (i,i):i\in V_{p}\}\cup \{(i,j):i<j\mbox{ and } (i,j)\in E\}.
$$
We denote by $M^{\nu(G)}$ the set of incomplete triangular matrices whose elements are indexed by $\nu(G)$ and whose diagonal elements are strictly positive. Both $\bfPhi^{\nu(G)}$ and $\bfPsi^{\nu(G)}$ must belong to $M^{\nu(G)}$. The non-free elements of $\bfPsi$ are determined through the completion operation from Lemma 2 of  \citet{atay-kayis_massam_2005} as a function of the free elements $\bfPsi^{\nu(G)}$. Each element $\Psi_{ij}$ with $i<j$ and $(i,j)\notin E$ is a function of the other elements $\Psi_{i^{\prime}j^{\prime}}$ that precede it in lexicographical order.\\
\indent \citet{roverato_2002} proves that the Jacobean of the transformation that maps $\bfK\in \cone_{G}$ to $\bfPhi^{\nu(G)}\in M^{\nu(G)}$ is
$$
 J\left(\bfK \rightarrow \bfPhi^{\nu(G)}\right) = 2^{p}\prod_{i=1}^{p}\Phi_{ii}^{v^{G}_{i}+1},
$$
where $v^{G}_{i}$ is the number of elements in the set $\{j:j>i \mbox{ and }(i,j)\in E\}$. Furthermore, \citet{atay-kayis_massam_2005} show  that the Jacobean of the transformation that maps $\bfPhi^{\nu(G)}\in M^{\nu(G)}$ to $\bfPsi^{\nu(G)}\in M^{\nu(G)}$ is given by
$$
  J\left(\bfPhi^{\nu(G)}\rightarrow \bfPsi^{\nu(G)}\right) = \prod_{i=1}^{p}Q_{ii}^{d^{G}_{i}+1},
$$
where $d^{G}_{i}$ is the number of elements in the set $\{j:j<i \mbox{ and }(i,j)\in E\}$. We have $\mbox{det}\; \bfK = \prod\limits_{i=1}^{p}\Phi_{ii}^{2}$ and $\Phi_{ii}=\Psi_{ii}Q_{ii}$. It follows that the density of $\bfPsi^{\nu(G)}$ with respect to the Lebesgue measure on $M^{\nu(G)}$ is
\begin{eqnarray*}
 p\left( \bfPsi^{\nu(G)}|\delta,\bfD\right) & = & p(\bfK | G,\delta,\bfD)J\left(\bfK \rightarrow \bfPhi^{\nu(G)}\right)J\left(\bfPhi^{\nu(G)}\rightarrow \bfPsi^{\nu(G)}\right),\\
 & = & \frac{2^{p}}{I_G(\delta,\bfD)} \prod\limits_{i=1}^{p}Q_{ii}^{v^{G}_{i}+d^{G}_{i}+\delta}\prod\limits_{i=1}^{p}\Psi_{ii}^{v^{G}_{i}+\delta-1}\exp\left(-\frac{1}{2}\sum\limits_{i,j=1}^{p}\Psi_{ij}^{2}\right).
\end{eqnarray*}
We note that $v^{G}_{i}+d^{G}_{i}$ represents the number of neighbors of vertex $i$ in the graph $G$.

\subsection{Sampling from the G-Wishart distribution} \label{sec:samplegwishart}

We introduce a Metropolis-Hastings algorithm for sampling from the G-Wishart distribution (\ref{eq:wishart}). We consider a strictly positive precision parameter $\sigma_{m}$. We denote by $\bfK^{[s]}=\bfQ^{T}\left( \bfPsi^{[s]}\right)^{T}\bfPsi^{[s]} \bfQ$ the current state of the chain with $ \left(\bfPsi^{[s]}\right)^{\nu(G)} \in M^{\nu(G)}$. The next state $\bfK^{[s+1]}=\bfQ^{T}\left( \bfPsi^{[s+1]}\right)^{T}\bfPsi^{[s+1]} \bfQ$ is obtained by sequentially perturbing the free elements $\left(\bfPsi^{[s]}\right)^{\nu(G)}$. A diagonal element $\Psi^{[s]}_{i_{0}i_{0}}>0$ is updated by sampling a value $\gamma$ from a $\normal\left( \Psi^{[s]}_{i_{0}i_{0}},\sigma_{m}^{2}\right)$ distribution truncated below at zero. We define the upper triangular matrix $\bfPsi^{\prime}$ such that $\Psi^{\prime}_{ij}=\Psi^{[s]}_{ij}$ for $(i,j)\in \nu(G)\setminus \{(i_{0},i_{0})\}$ and $\Psi^{\prime}_{i_{0}i_{0}}=\gamma$. The non-free elements of $\bfPsi^{\prime}$ are obtained through the completion operation from $\left(\bfPsi^{\prime}\right)^{\nu(G)}$. The Markov chain moves to $\bfK^{\prime}=\bfQ^{T}\left( \bfPsi^{\prime}\right)^{T}\bfPsi^{\prime} \bfQ$ with probability $\min\left\{ R_{m},1\right\}$, where
\begin{eqnarray} \label{eq:rprime}
 R_{m} = \frac{p\left( \left(\bfPsi^{\prime}\right)^{\nu(G)}|\delta,\bfD\right) }{p\left( \left(\bfPsi^{[s]}\right)^{\nu(G)}|\delta,\bfD\right)}\frac{p\left(\Psi^{[s]}_{i_{0}i_{0}}|\Psi^{\prime}_{i_{0}i_{0}}\right) }{p\left(\Psi^{\prime}_{i_{0}i_{0}}|\Psi^{[s]}_{i_{0}i_{0}}\right) } = \frac{\phi\left(\Psi^{[s]}_{i_{0}i_{0}}/\sigma_{m}\right)}{\phi\left(\Psi^{\prime}_{i_{0}i_{0}}/\sigma_{m}\right)} \left( \frac{\Psi^{\prime}_{i_{0}i_{0}}}{\Psi^{[s]}_{i_{0}i_{0}}} \right)^{v^{G}_{i_{0}}+\delta-1} R^{\prime}_{m}.
 \end{eqnarray}
Here $\phi(\cdot)$ represents the CDF of the standard normal distribution and
\begin{eqnarray}\label{eq:rprime}
R^{\prime}_{m} =\exp\left\{-\frac{1}{2} \sum\limits_{i,j=1}^{p} \left[ \left(\Psi^{\prime}_{ij}\right)^{2}-\left(\Psi^{[s]}_{ij}\right)^{2}\right]\right\}.
\end{eqnarray}
\noindent A free off-diagonal element $\Psi^{[s]}_{i_{0}j_{0}}$ is updated by sampling a value $\gamma^{\prime}\sim \normal\left( \Psi^{[s]}_{i_{0}j_{0}},\sigma_{m}^{2}\right)$. We define the upper triangular matrix $\bfPsi^{\prime}$ such that $\Psi^{\prime}_{ij}=\Psi^{[s]}_{ij}$ for $(i,j)\in \nu(G)\setminus \{(i_{0},j_{0})\}$ and $\Psi^{\prime}_{i_{0}j_{0}}=\gamma^{\prime}$. The remaining elements of $\bfPsi^{\prime}$ are determined by completion from $\left(\bfPsi^{\prime}\right)^{\nu(G)}$. The proposal distribution is symmetric $p\left(\Psi^{\prime}_{i_{0}j_{0}}| \Psi^{[s]}_{i_{0}j_{0}}\right)=p\left( \Psi^{[s]}_{i_{0}j_{0}}|\Psi^{\prime}_{i_{0}j_{0}}\right)$, thus we accept the transition of the chain from $\bfK^{[s]}$ to $\bfK^{\prime}=\bfQ^{T}\left( \bfPsi^{\prime}\right)^{T}\bfPsi^{\prime} \bfQ$ with probability $\min\{ R^{\prime}_{m},1\}$, where $R^{\prime}_{m}$ is given in equation (\ref{eq:rprime}). Since $ \left(\bfPsi^{[s]}\right)^{\nu(G)} \in M^{\nu(G)}$, we have $ \left(\bfPsi^{\prime}\right)^{\nu(G)} \in M^{\nu(G)}$ which implies $\bfK^{\prime}\in \cone_{G}$.\\
\indent We denote by $\bfK^{[s+1]}$ the precision matrix obtained after completing all the Metropolis-Hastings updates associated with the free elements indexed by $\nu(G)$. The acceptance probabilities depend only on the free elements of the upper diagonal matrices $\bfPsi^{[s]}$ and  $\bfPsi^{\prime}$. An efficient implementation of our sampling procedure calculates only the final matrix $\bfK^{[s+1]}$ and avoids determining the intermediate candidate precision matrices $\bfK^{\prime}$. There is another key computational aspect that relates to the dependence of the Cholesky decomposition on a particular ordering of the variables. Consider two free elements $\Psi^{[s]}_{i_{0}j_{0}}$ and $\Psi^{[s]}_{i^{\prime}_{0}j^{\prime}_{0}}$ such that $(i_{0},j_{0})<(i_{0}^{\prime},j_{0}^{\prime})$ in lexicographical order. The completion operation from Lemma 2 of  \citet{atay-kayis_massam_2005} shows that perturbing the value of $\Psi^{[s]}_{i_{0}j_{0}}$ leads to a change in the value of all non-free elements $\left\{ \Psi^{[s]}_{ij}:(i,j)\notin E \mbox{ and } (i_{0},j_{0})<(i,j)\right\}$. On the other hand, perturbing the value of $\Psi^{[s]}_{i^{\prime}_{0}j^{\prime}_{0}}$ implies a change in the smaller set of elements $\left\{ \Psi^{[s]}_{ij}:(i,j)\notin E \mbox{ and } (i^{\prime}_{0},j^{\prime}_{0})<(i,j)\right\}$. This means that the chain is less likely to move when performing the update associated with $\Psi^{[s]}_{i_{0}j_{0}}$ than when updating $\Psi^{[s]}_{i^{\prime}_{0}j^{\prime}_{0}}$. As such, the ordering of the variables is modified at each iteration of the MCMC sampler. More specifically, a permutation $\upsilon$ is uniformly drawn from the set of all possible permutations $\Upsilon_{p}$ of $V$. The row and columns of $\bfD$ are reordered according to $\upsilon$ and a new Cholesky decomposition of $\bfD^{-1}$ is determined. The set $\nu(G)$ and $\{d_{i}^{G}:i\in V\}$ are recalculated given the ordering of the vertices induced by $\upsilon$, i.e. $i<_{\upsilon}j$ if and only if $\upsilon(i)<\upsilon(j)$. Changing the ordering guarantees that across iterations it is equally likely that $(i_{0},j_{0})$ precedes or succeeds $(i_{0}^{\prime},j_{0}^{\prime})$.\\
\indent Our later developments from Section \ref{sec:matrixggm} involve sampling from the G-Wishart distribution (\ref{eq:wishart}) subject to the constraint $K_{11}=1$. We have
$$
 K^{[s]}_{11}=1 \Leftrightarrow \left( \bfPsi^{[s]} \bfQ \right)_{11} = 1 \Leftrightarrow \Psi^{[s]}_{11} = 1/Q_{11}.
$$
We subsequently obtain the next state $\bfK^{[s+1]}$ of the Markov chain by perturbing the free elements $\left(\bfPsi^{[s]}\right)^{\nu(G)\setminus \{(1,1)\}}$. When defining the triangular matrix $\bfPsi^{\prime}$ we set $\Psi^{\prime}_{11} = 1/Q_{11}$ which implies that the corresponding candidate matrix $\bfK^{\prime}$ has $K^{\prime}_{11}=1$. Thus $\bfK^{[s+1]}$ also obeys the constraint $K^{[s+1]}_{11}=1$. The random orderings of the variables need to be drawn from the set $\Upsilon_{p}^{(1,1)}$ of permutations $\upsilon\in \Upsilon_{p}$ such that $\upsilon(1)=1$. This way the $(1,1)$ element of $\bfK$ always occupies the same position.

\subsection{Bayesian inference in GGMs} \label{sec:graphmcmc}

We let $\mathcal{D}=\left\{\bfx^{(1)},\ldots,\bfx^{(n)}\right\}$ be the observed data of $n$ independent
samples from $\normal_{p}(\mathbf{0},\bfK^{-1})$. The likelihood function is
\begin{eqnarray} \label{eq:likelihood}
  p\left(\mathcal{D}|\bfK\right) = (2\pi)^{-np/2} (\mbox{det}\; \bfK)^{n/2}\exp\left\{-\frac{1}{2}\langle \bfK,\bfU \rangle\right\},
\end{eqnarray}
\noindent where $\bfU = \sum\limits_{j=1}^n\bfx^{(j)}(\bfx^{(j)})^{T}$ is the observed sum-of-products matrix. Given a graph $G$, we assume a G-Wishart prior $\GWis_{G}(\delta_{0},\bfD_{0})$ for the precision matrix $\bfK$. We take $\delta_{0}=3>2$ and $D_{0}=\mathbf{I}_{p}$, the $p$-dimensional identity matrix. This choice implies that prior for $\bfK$ is equivalent with one observed sample,  while the observed variables are assumed to be apriori independent of each other. Since the G-Wishart prior for $\bfK$ is conjugate for the likelihood (\ref{eq:likelihood}), the posterior of $\bfK$ given $G$ is $\GWis_{G}(n+\delta_{0},\bfU+\bfD_{0})$. We also assume a uniform prior $p(G) = 2/p(p-1)$ on $\mathcal{G}_{p}$.\\
\indent We develop a MCMC algorithm for sampling from the joint distribution
$$
 p\left( \mathcal{D},\bfK,G|\delta_{0},\bfD_{0}\right) \propto p\left( \mathcal{D}|\bfK\right)p\left(\bfK|G,\delta_{0},\bfD_{0}\right)p(G),
$$
that is well-defined if and only if $\bfK\in \cone_{G}$. We denote the current state of the chain by $(\bfK^{[s]},G^{[s]})$, $K^{[s]}\in \cone_{G^{[s]}}$. Its next state $(\bfK^{[s+1]},G^{[s+1]})$, $\bfK^{[s+1]}\in \cone_{G^{[s+1]}}$, is generated by sequentially performing the following two steps. We make use of two strictly positive precision parameters $\sigma_{m}$ and $\sigma_{g}$ that remain fixed at some suitable small values. We assume that the ordering of the variables has been changed according to a permutation $\upsilon$ selected at random from the uniform distribution on $\Upsilon_{p}$. We denote by $(\bfU+\bfD_{0})^{-1}=(\bfQ^{*})^{T}\bfQ^{*}$ the Cholesky decomposition of $(\bfU+\bfD_{0})^{-1}$, where the rows and columns of this matrix have been permuted according to $\upsilon$.\\
\indent  We denote by $\nbd_{p}^{+}(G)$ the graphs that can be obtained by adding an edge to a graph $G\in \ground_{p}$ and by $\nbd_{p}^{-}(G)$ the graphs that are obtained by deleting an edge from $G$. We call the one-edge-way set of graphs $\nbd_{p}(G)=\nbd_{p}^{+}(G)\bigcup \nbd_{p}^{-}(G)$ the neighborhood of $G$ in $\ground_{p}$. These neighborhoods connect any two graphs in $\ground_{p}$ through a sequence of graphs such that two consecutive graphs in this sequence are each others' neighbors.

$ $\\
{\bf Step 1: Resample the graph}. We sample a candidate graph $G^{\prime}\in \nbd_{p}\left(G^{[s]}\right)$ from the proposal
\begin{eqnarray} \label{eq:fairproposal}
 q\left(G^{\prime}|G^{[s]}\right) = \frac{1}{2}\Uni\left( \nbd_{p}^{+}\left(G^{[s]}\right)\right) + \frac{1}{2}\Uni\left( \nbd_{p}^{-}\left(G^{[s]}\right)\right),
\end{eqnarray}
\noindent where $\Uni(A)$ represents the uniform distribution on the discrete set $A$. The distribution (\ref{eq:fairproposal}) gives an equal probability of proposing to delete an edge from the current graph and of proposing to add an edge to the current graph. We favor (\ref{eq:fairproposal}) over the more usual proposal distribution $\Uni\left( \nbd_{p}\left(G^{[s]}\right)\right)$ that is employed, for example, by \citet{madigan_york_1995}. If $G^{[s]}$ contains a very large or a very small number of edges, the probability of proposing a move that adds or, respectively, deletes an edge from $G^{[s]}$ is extremely small when sampling from $\Uni\left( \nbd_{p}\left(G^{[s]}\right)\right)$, which could lead to poor mixing in the resulting Markov chain.\\
\indent We assume that the candidate $G^{\prime}$ sampled from (\ref{eq:fairproposal}) is obtained by adding the edge  $(i_{0},j_{0})$, $i_{0}<j_{0}$, to $G^{[s]}$. Since $G^{\prime}\in \nbd_{p}^{+}\left(G^{[s]}\right)$ we have $G^{[s]}\in \nbd_{p}^{-}\left(G^{\prime}\right)$. We consider the decomposition of the current precision matrix
$$
 \bfK^{[s]} = (\bfQ^{*})^{T}\left( (\bfPsi^{[s]})^{T}\bfPsi^{[s]}\right) \bfQ^{*},
$$
\noindent with $(\bfPsi^{[s]})^{\nu(G^{[s]})}\in M^{\nu(G^{[s]})}$. Since the vertex $i_{0}$ has one additional neighbor in $G^{\prime}$, we have $d_{i_{0}}^{G^{\prime}}=d_{i_{0}}^{G^{[s]}}$, $d_{j_{0}}^{G^{\prime}}=d_{j_{0}}^{G^{[s]}}+1$, $v_{i_{0}}^{G^{\prime}}=v_{i_{0}}^{G^{[s]}}+1$, $v_{j_{0}}^{G^{\prime}}=v_{j_{0}}^{G^{[s]}}$ and $\nu\left( G^{\prime}\right) = \nu\left( G^{[s]}\right)\cup \{(i_{0},j_{0})\}$. We define an upper triangular matrix $\bfPsi^{\prime}$ such that $\Psi^{\prime}_{ij}=\Psi^{[s]}_{ij}$ for $(i,j)\in \nu(G^{[s]})$. We sample $\gamma \sim \normal\left( \Psi^{[s]}_{i_{0}j_{0}},\sigma^{2}_{g}\right)$ and set $\Psi^{\prime}_{i_{0}j_{0}}=\gamma$. The rest of the elements of $\bfPsi^{\prime}$ are determined from $(\bfPsi^{\prime})^{\nu(G^{\prime})}$ through the completion operation. The value of the free element $\Psi^{\prime}_{i_{0}j_{0}}$ was set by perturbing the non-free element $\Psi^{[s]}_{i_{0}j_{0}}$. The other free elements of $\bfPsi^{\prime}$ and $\bfPsi^{[s]}$ coincide.\\
\indent We take $\bfK^{\prime} = (\bfQ^{*})^{T}\left( (\bfPsi^{\prime})^{T}\bfPsi^{\prime}\right) \bfQ^{*}$. Since $(\bfPsi^{\prime})^{\nu(G^{\prime})}\in M^{\nu(G^{\prime})}$, we have $\bfK^{\prime} \in \cone_{G^{\prime}}$. The dimensionality of the parameter space increases by one as we move from $(\bfK^{[s]},G^{[s]})$ to $(\bfK^{\prime},G^{\prime})$, thus we must make use of the reversible jump algorithm of \citet{Gr95}. The Markov chain moves to $(\bfK^{\prime},G^{\prime})$ with probability $\min\left\{ R^{+}_{g},1\right\}$ where $R^{+}_{g}$ is given by
\begin{eqnarray} \label{eq:rg}
 R^{+}_{g} & = & \frac{p\left(\mathcal{D}|\bfK^{\prime}\right) }{p\left(\mathcal{D}|\bfK^{[s]}\right)} \frac{ p\left( \bfK^{\prime}|G^{\prime},\delta_{0},\bfD_{0} \right) }{ p\left( \bfK^{[s]}|G^{[s]},\delta_{0},\bfD_{0} \right) } \frac{| \nbd_{p}^{+}\left(G^{[s]}\right)|}{| \nbd_{p}^{-}\left(G^{\prime}\right)|}\times\\
 &&  \times \frac{J\left(\bfK^{\prime} \rightarrow (\bfPsi^{\prime})^{\nu(G^{\prime})}\right)}{J\left(\bfK^{[s]} \rightarrow (\bfPsi^{[s]})^{\nu(G^{[s]})}\right)}\frac{J\left( \left( (\bfPsi^{[s]})^{\nu(G^{[s]})}, \gamma \right) \rightarrow (\bfPsi^{\prime})^{\nu(G^{\prime})}\right)}{\frac{1}{\sigma_{g}\sqrt{2\pi}}\exp\left( -\frac{\left(\Psi^{\prime}_{i_{0}j_{0}}-\Psi^{[s]}_{i_{0}j_{0}}\right)^{2}}{2\sigma_{g}^{2}}\right)},\nonumber
\end{eqnarray}
\noindent where $|A|$ denotes the number of elements of the set $A$. Otherwise the chain stays at $(\bfK^{[s]},G^{[s]})$. Since $(\bfPsi^{\prime})^{\nu(G^{[s]})} = (\bfPsi^{[s]})^{\nu(G^{[s]})}$, the Jacobean of the transformation from $\left( (\bfPsi^{[s]})^{\nu(G^{[s]})}, \gamma \right)$  to $(\bfPsi^{\prime})^{\nu(G^{\prime})}$ is equal to $1$. Moreover, $\bfPsi^{[s]}$ and $\bfPsi^{\prime}$ have the same diagonal elements, hence $\mbox{det}\; \bfK^{[s]}=\prod\limits_{i=1}^{p}\left(Q^{*}_{ii}\Psi^{[s]}_{ii}\right)^{2}=\mbox{det}\; \bfK^{\prime}$. It follows that equation (\ref{eq:rg}) becomes
\begin{eqnarray} \label{eq:rgplus}
 R^{+}_{g} & = & \sigma_{g}\sqrt{2\pi}Q^{*}_{i_{0}i_{0}}Q^{*}_{j_{0}j_{0}}\Psi^{[s]}_{i_{0}i_{0}}\frac{I_{G^{[s]}}(\delta_{0},\bfD_{0})}{I_{G^{\prime}}(\delta_{0},\bfD_{0})}  \frac{| \nbd_{p}^{+}\left(G^{[s]}\right)|}{| \nbd_{p}^{-}\left(G^{\prime}\right)|}\times\\
 && \times \exp\left\{-\frac{1}{2}\left[\left\langle \bfK^{\prime}-\bfK^{[s]},\bfU+\bfD_{0}\right\rangle -\frac{\left(\Psi^{\prime}_{i_{0}j_{0}}-\Psi^{[s]}_{i_{0}j_{0}}\right)^{2}}{\sigma_{g}^{2}}\right]\right\}.\nonumber
\end{eqnarray}
\indent Next we assume that the candidate $G^{\prime}$ is obtained by deleting the edge $(i_{0},j_{0})$ from $G^{[s]}$. We have $d_{i_{0}}^{G^{\prime}}=d_{i_{0}}^{G^{[s]}}$, $d_{j_{0}}^{G^{\prime}}=d_{j_{0}}^{G^{[s]}}-1$, $v_{i_{0}}^{G^{\prime}}=v_{i_{0}}^{G^{[s]}}-1$, $v_{j_{0}}^{G^{\prime}}=v_{j_{0}}^{G^{[s]}}$ and $\nu\left( G^{\prime}\right) = \nu\left( G^{[s]}\right)\setminus \{(i_{0},j_{0})\}$. We define an upper triangular matrix $\bfPsi^{\prime}$ such that $\Psi^{\prime}_{ij}=\Psi^{[s]}_{ij}$ for $(i,j)\in \nu(G^{\prime})$. The rest of the elements of $\bfPsi^{\prime}$ are determined through completion. The free element $\Psi^{[s]}_{i_{0}j_{0}}$ becomes non-free in $\bfPsi^{\prime}$, hence the parameter space decreases by 1 as we move from  $(\bfPsi^{[s]})^{\nu(G^{[s]})}$ to $(\bfPsi^{\prime})^{\nu(G^{\prime})}\in M^{\nu(G^{\prime})}$. As before, we take $\bfK^{\prime} = (\bfQ^{*})^{T}\left( (\bfPsi^{\prime})^{T}\bfPsi^{\prime}\right) \bfQ^{*}$. The acceptance probability of the transition from $(\bfK^{[s]},G^{[s]})$ to $(\bfK^{\prime},G^{\prime})$ is $\min\left\{ R^{-}_{g},1\right\}$ where
\begin{eqnarray} \label{eq:rgminus}
 R^{-}_{g}& = & \left(\sigma_{g}\sqrt{2\pi}Q^{*}_{i_{0}i_{0}}Q^{*}_{j_{0}j_{0}}\Psi^{[s]}_{i_{0}i_{0}}\right)^{-1}\frac{I_{G^{[s]}}(\delta_{0},\bfD_{0})}{I_{G^{\prime}}(\delta_{0},\bfD_{0})} \frac{| \nbd_{p}^{-}\left(G^{[s]}\right)|}{| \nbd_{p}^{+}\left(G^{\prime}\right)|}\times\\
 && \times\exp\left\{-\frac{1}{2}\left[\left\langle \bfK^{\prime}-\bfK^{[s]},\bfU+\bfD_{0}\right\rangle +\frac{\left(\Psi^{\prime}_{i_{0}j_{0}}-\Psi^{[s]}_{i_{0}j_{0}}\right)^{2}}{\sigma_{g}^{2}}\right]\right\}.\nonumber
\end{eqnarray}
\noindent We denote by $(\bfK^{[s+1/2]},G^{[s+1]})$, $\bfK^{[s+1/2]}\in G^{[s+1]}$, the state of the chain at the end of this step. 

$ $\\
{\bf Step 2: Resample the precision matrix}. Given the updated graph $G^{[s+1]}$, we update the precision matrix $\bfK^{[s+1/2]}=(\bfQ^{*})^{T}\left( \bfPsi^{s+1/2}\right)^{T}\bfPsi^{s+1/2} \bfQ^{*}$ by sequentially perturbing the free elements $\left(\bfPsi^{[s+1/2]}\right)^{\nu\left(G^{[s+1]}\right)} $. For each such element, we perform the corresponding Metropolis-Hastings step from Section \ref{sec:graphmcmc}  with $\delta = n+\delta_{0}$, $\bfD = \bfU+\bfD_{0}$ and $\bfQ = \bfQ^{*}$. The standard deviation of the normal proposals is $\sigma_{m}$. We denote by $\bfK^{[s+1]}\in \cone_{G^{[s+1]}}$ the precision matrix obtained after all the updates have been performed.

\section{Matrix-variate Gaussian graphical models} \label{sec:matrixggm}

We extend our framework to the case when the observed data $\mathcal{D}=\left\{\bfx^{(1)},\ldots,\bfx^{(n)}\right\}$ are associated with a $p_{R}\times p_{C}$ random matrix $\bfX=(X_{ij})$ that follows a matrix-variate normal distribution
$$
 \mbox{vec}\left(\bfX^{T}\right)|\bfK_{R},\bfK_{C}\sim \normal_{p_{R}p_{C}}\left(\bf0,\left[\bfK_{R}\otimes \bfK_{C}\right]^{-1}\right)
$$
\noindent with p.d.f. \citep{gupta_nagar_2000}:
\begin{eqnarray} \label{eq:matrixnormal}
 p\left(\bfX|\bfK_{R},\bfK_{C}\right) = (2\pi)^{-p_{R}p_{C}/2} \left(\mbox{det}\; \bfK_{R}\right)^{p_{C}/2}\left(\mbox{det}\; \bfK_{C}\right)^{p_{R}/2}\exp\left\{-\frac{1}{2}\mbox{tr}\left[  \bfK_{R}\bfX\bfK_{C}\bfX^{T}\right]\right\}.
\end{eqnarray}
\noindent Here $\bfK_{R}$ is a $p_{R}\times p_{R}$ row precision matrix and $\bfK_{C}$ is a $p_{C}\times p_{C}$ column precision matrix. Furthermore, we assume that $\bfK_{R}\in \cone_{G_{R}}$ and $\bfK_{C}\in \cone_{G_{C}}$ where $G_{R}=\left(V_{p_{R}},E_{R}\right)$ and $G_{C}=\left(V_{p_{C}},E_{C}\right)$ are two graphs with $p_{R}$ and $p_{C}$ vertices, respectively. We consider the rows $\bfX_{1\textasteriskcentered},\ldots,\bfX_{p_{R}\textasteriskcentered}$ and the columns $\bfX_{\textasteriskcentered 1},\ldots,\bfX_{\textasteriskcentered p_{C}}$ of the random matrix $\bfX$. From Theorem 2.3.12 of \citet{gupta_nagar_2000} we have $\bfX_{i\textasteriskcentered}^{T}\sim \normal_{p_{C}}\left(\mathbf{0},\left(\bfK^{-1}_{R}\right)_{ii}\bfK^{-1}_{C}\right)$ and $\bfX_{\textasteriskcentered j}\sim \normal_{p_{R}}\left(\mathbf{0},\left(\bfK^{-1}_{C}\right)_{jj}\bfK^{-1}_{R}\right)$. The graphs $G_{R}$ and $G_{C}$ define GGMs for the rows and columns of $\bfX$ \citep{wang_west_2009}:
\begin{eqnarray}\label{eq:matvarci}
 \bfX_{i_1\textasteriskcentered} \Perp \bfX_{i_2\textasteriskcentered}\mid \bfX_{(V_{p_{R}}\setminus \{i_1,i_2\})\textasteriskcentered} & \Leftrightarrow & \left(\bfK_{R}\right)_{i_1i_2}=\left(\bfK_{R}\right)_{i_2i_1}=0 \Leftrightarrow (i_{1},i_{2})\notin E_{R}, \mbox{ and }\\
 \bfX_{\textasteriskcentered j_{1}} \Perp \bfX_{\textasteriskcentered j_{2}}\mid \bfX_{(V_{p_{C}}\setminus \{j_1,j_2\})\textasteriskcentered} & \Leftrightarrow & \left(\bfK_{C}\right)_{j_1j_2}=\left(\bfK_{R}\right)_{j_2j_1}=0 \Leftrightarrow (j_{1},j_{2})\notin E_{C}.\nonumber
\end{eqnarray}
Any prior specification for $\bfK_{R}$ and $\bfK_{C}$ must take into account the fact that the two precision matrices are not uniquely identified from their Kronecker product which means that, for any $z>0$, $\left(z^{-1}\bfK_{R}\right)\otimes \left(z\bfK_{C}\right) = \bfK_{R}\otimes \bfK_{C}$ represents the same precision matrix for $\mbox{vec}(\bfX^{T})$ \--- see equation (\ref{eq:matrixnormal}). We follow the basic idea laid out in \citet{wang_west_2009} and impose the constraint $\left(\bfK_{C}\right)_{11}=1$. Furthermore, we define a prior for $\bfK_{C}$ through parameter expansion by assuming a G-Wishart prior $\GWis_{G_{C}}(\delta_{C},\bfD_{C})$ for the matrix $z \bfK_{C}$ with $z>0$, $\delta_{C}>2$ and $D_{C}\in \cone_{G_{C}}$. It is immediate to see that the Jacobean of the transformation from $z\bfK_{C}$ to $\left( z, \bfK_{C}\right)$ is 
$$
 J\left( \left( z\bfK_{C} \right) \rightarrow \left( z, \bfK_{C}\right) \right) = z^{|\nu\left( G_{C}\right)|-1}.
$$
It follows that our joint prior for $\left( z, \bfK_{C}\right)$ is given by
$$
 p\left( z, \bfK_{C}| G_{C},\delta_{C},\bfD_{C}\right) = \frac{1}{I_{G_{C}}\left(\delta_{C},\bfD_{C}\right)}\left(\mbox{det}\; \bfK_{C}\right)^{\frac{\delta_{C}-2}{2}}\exp\left\{-\frac{1}{2}\langle \bfK_{C},z\bfD_{C}\rangle\right\} z^{\frac{p_{C}(\delta_{C}-2)}{2}+|\nu\left( G_{C}\right)|-1}.
$$
The elements of $\bfK_{R}\in \cone_{G_{R}}$ are not subject to any additional constraints, hence we assume a G-Wishart prior $\GWis_{G_{R}}(\delta_{R},\bfD_{R})$ for $\bfK_{R}$. We take $\delta_{C}=\delta_{R}=3$, $\bfD_{C}=\mathbf{I}_{p_{C}}$ and $\bfD_{R}=\mathbf{I}_{p_{R}}$. We complete our prior specification by assuming uniform priors $p\left( G_{C}\right) = 2/p_{C}(p_{C}-1)$ and $p\left( G_{R}\right) = 2/p_{R}(p_{R}-1)$ for the row and column graphs, where $G_{R}\in \mathcal{G}_{p_{R}}$ and $G_{C}\in \mathcal{G}_{p_{C}}$.\\
\indent We perform Bayesian inference for matrix-variate GGMs by developing a MCMC algorithm for sampling from the joint distribution of the data, row and column precision matrices and row and column graphs
\begin{eqnarray}\label{eq:matrixjoint}
p\left( \mathcal{D}, \bfK_{R},z,\bfK_{C},G_{R},G_{C}|\delta_{C},\bfD_{C},\delta_{R},\bfD_{R}\right) & \propto & p\left( \mathcal{D}|\bfK_{R},\bfK_{C}\right)\times \\
 && \hspace{-2cm}\times p\left( \bfK_{R}| G_{R},\delta_{R},\bfD_{R}\right) p\left( z, \bfK_{C}| G_{C},\delta_{C},\bfD_{C}\right) p\left( G_{C}\right) p\left( G_{R}\right),\nonumber
\end{eqnarray}
that is well-defined for $\bfK_{R}\in \cone_{G_{R}}$, $\bfK_{C}\in \cone_{G_{C}}$ with $\left(\bfK_{C}\right)_{11}=1$ and $z>0$. Equation (\ref{eq:matrixjoint}) is written as:
\begin{eqnarray} \label{eq:matrixjoint2}
 p\left( \mathcal{D}, \bfK_{R},z,\bfK_{C},G_{R},G_{C}|\delta_{C},\bfD_{C},\delta_{R},\bfD_{R}\right) \propto z^{\frac{p_{C}(\delta_{C}-2)}{2}+|\nu\left( G_{C}\right)|-1} \left(\mbox{det}\; \bfK_{R}\right)^{\frac{np_{C}+\delta_{R}-2}{2}}\left(\mbox{det}\; \bfK_{C}\right)^{\frac{np_{R}+\delta_{C}-2}{2}}\times && \\
 && \hspace{-9cm}\times \exp\left\{ -\frac{1}{2}\mbox{tr}\left[ \sum\limits_{j=1}^{n} \bfK_{R}\bfx^{(j)}\bfK_{C}\left( \bfx^{(j)}\right)^{T}+\bfK_{R}\bfD_{R}+\bfK_{C}\left(z\bfD_{C}\right)\right]\right\}.\nonumber
\end{eqnarray}
Our sampling scheme is comprised of the following five steps that explain the transition of the Markov chain from its current state $(\bfK_{R}^{[s]},\bfK_{C}^{[s]},G_{R}^{[s]},z^{[s]})$ to its next state $(\bfK_{R}^{[s+1]},\bfK_{C}^{[s+1]},G_{R}^{[s+1]},z^{[s+1]})$. We use four strictly positive precision parameters $\sigma_{m,R}$, $\sigma_{m,C}$, $\sigma_{g,R}$ and $\sigma_{g,C}$.\\
$ $\\
{\bf Step 1: Resample the row graph}. We denote $n^{*}_{R}=np_{C}$ and $\bfU^{*}_{R}=\sum\limits_{j=1}^{n}\bfx^{(j)}\bfK_{C}^{[s]}\left( \bfx^{(j)}\right)^{T}$. We generate a random permutation $\upsilon_{R}\in \Upsilon_{p_{R}}$ of the row indices $V_{p_{R}}$ and reorder the row and columns of the matrix $\bfU^{*}_{R}+\bfD_{R}$ according to $\upsilon_{R}$. We determine the Cholesky decomposition $\left(\bfU^{*}_{R}+\bfD_{R}\right)^{-1}=\left( \bfQ^{*}_{R}\right)^{T}\bfQ^{*}_{R}$. We proceed as described in Step 1 of Section \ref{sec:graphmcmc}. Given the notations we used in that section, we take $p=p_{R}$, $n = n^{*}_{R}$, $\bfU = \bfU^{*}_{R}$, $\delta_{0}=\delta_{R}$, $\bfD_{0}=\bfD_{R}$ and $\sigma_{g}=\sigma_{g,R}$. We denote the updated row precision matrix and graph by $\left( \bfK_{R}^{[s+1/2]},G_{R}^{[s+1]}\right)$, $\bfK_{R}^{[s+1/2]}\in \cone_{G_{R}^{[s+1]}}$.\\
$ $\\
{\bf Step 2: Resample the row precision matrix}. We denote $n^{*}_{R}=np_{C}$ and $\bfU^{*}_{R}=\sum\limits_{j=1}^{n}\bfx^{(j)}\bfK_{C}^{[s]}\left( \bfx^{(j)}\right)^{T}$. We determine the Cholesky decomposition $\left(\bfU^{*}_{R}+\bfD_{R}\right)^{-1}=\left( \bfQ^{*}_{R}\right)^{T}\bfQ^{*}_{R}$ after permuting the row and columns of $\bfU^{*}_{R}+\bfD_{R}$ according to a random ordering in $\Upsilon_{p_{R}}$.
The conditional distribution of $\bfK_{R}\in \cone_{G_{R}^{[s+1]}}$ is G-Wishart  $\GWis_{G_{R}^{[s+1]}}\left( n^{*}_{R}+\delta_{R},   \bfU^{*}_{R}  + \bfD_{R}\right)$. We make the transition from $\bfK_{R}^{[s+1/2]}$ to $\bfK_{R}^{[s+1]}\in \cone_{G_{R}^{[s+1]}}$ using Metropolis-Hastings updates described in Section \ref{sec:samplegwishart}. Given the notations we used in that section, we take $p=p_{R}$,  $\delta = n^{*}_{R}+\delta_{R}$, $\bfD = \bfU^{*}_{R}  + \bfD_{R}$, $\bfQ = \bfQ^{*}_{R}$ and $\sigma_{m}=\sigma_{m,R}$.\\
$ $\\
{\bf Step 3: Resample the column graph}. We denote $n^{*}_{C}=np_{R}$ and $\bfU^{*}_{C}=\sum\limits_{j=1}^{n}\left( \bfx^{(j)}\right)^{T}\bfK_{R}^{[s+1]}\bfx^{(j)}$. The relevant conditional distribution is (see equation (\ref{eq:matrixjoint2})):
\begin{eqnarray}\label{eq:rcg}
 p\left(\bfK_{C}^{[s]},G_{C}^{[s]},z^{[s]}|\mathcal{D},\bfK^{[s+1]}_{R},\delta_{C},\bfD_{C}\right) & \propto & \frac{1}{I_{G_{C}^{[s]}}(\delta_{C},\bfD_{C})} \left(z^{[s]}\right)^{|\nu\left( G_{C}\right)|} \left(\mbox{det}\; \bfK_{C}^{[s]}\right)^{\frac{n^{*}_{C}+\delta_{C}-2}{2}} \times\\
  && \times \exp\left\{ -\frac{1}{2} \left\langle \bfK_{C}^{[s]},\bfU^{*}_{C} +z^{[s]}\bfD_{C}\right\rangle\right\}.\nonumber
\end{eqnarray}
We sample a candidate column graph $G_{C}^{\prime}\in \nbd_{p_{C}}\left(G_{C}^{[s]}\right)$ from the proposal
\begin{eqnarray} \label{eq:fairproposalcol}
 q\left( G_{C}^{\prime}|G_{C}^{[s]},z^{[s]}\right) = \frac{1}{2}\frac{\left(z^{[s]}\right)^{|\nu\left( G^{\prime}_{C}\right)|}}{\sum\limits_{G^{\prime\prime}_{C}\in \nbd^{+}_{p_{C}}\left(G_{C}^{[s]}\right)}\left( z^{[s]} \right)^{|\nu\left( G^{\prime\prime}_{C}\right)|}} \mathbf{1}_{\left\{ G^{\prime}_{C}\in \nbd^{+}_{p_{C}}\left(G_{C}^{[s]}\right)\right\}} + \frac{1}{2}\frac{\left(z^{[s]}\right)^{|\nu\left( G^{\prime}_{C}\right)|}}{\sum\limits_{G^{\prime\prime}_{C}\in \nbd^{-}_{p_{C}}\left(G_{C}^{[s]}\right)}\left( z^{[s]} \right)^{|\nu\left( G^{\prime\prime}_{C}\right)|}} \mathbf{1}_{\left\{ G^{\prime}_{C}\in \nbd^{-}_{p_{C}}\left(G_{C}^{[s]}\right)\right\}},
\end{eqnarray}
\noindent where $1_{A}$ is equal to $1$ if $A$ is true and is $0$ otherwise. The proposal (\ref{eq:fairproposalcol}) gives an equal probability that the candidate graph is obtained by adding or deleting an edge from the current graph \--- see also equation (\ref{eq:fairproposal}).\\
\indent We assume that $G_{C}^{\prime}$ is obtained by adding the edge $(i_{0},j_{0})$ to $G_{C}^{[s]}$. We generate a random permutation $\upsilon_{C}\in \Upsilon_{p_{C}}^{(1,1)}$ of the row indices $V_{p_{C}}$ and reorder the row and columns of the matrix $\bfU^{*}_{C}+z^{[s]}\bfD_{C}$ according to $\upsilon_{C}$. The permutation $\upsilon_{C}$ is such that $\upsilon_{C}(1)=1$, hence the $(1,1)$ element of $\bfK_{C}^{[s]}$ remains in the same position. We determine the Cholesky decomposition $\left(\bfU^{*}_{C}+z^{[s]}\bfD_{C}\right)^{-1}=\left( \bfQ^{*}_{C}\right)^{T}\bfQ^{*}_{C}$ of $\left(\bfU^{*}_{C}+z^{[s]}\bfD_{C}\right)^{-1}$. We consider the decomposition of the column precision matrix
$$
 \bfK_{C}^{[s]} = \left(\bfQ^{*}_{C}\right)^{T}\left(\bfPsi_{C}^{[s]}\right)^{T}\bfPsi_{C}^{[s]} \bfQ^{*}_{C},
$$
\noindent with $\left(\bfPsi_{C}^{[s]}\right)^{\nu\left(G_{C}^{[s]}\right)}\in M^{\nu\left(G_{C}^{[s]}\right)}$. We define an upper triangular matrix $\bfPsi^{\prime}_{C}$ such that $\left(\bfPsi^{\prime}_{C}\right)_{ij}=\left(\bfPsi^{[s]}_{C}\right)_{ij}$ for $(i,j)\in \nu\left(G^{[s]}_{C}\right)$. We sample $\gamma \sim \normal\left( \left(\bfPsi^{[s]}_{C}\right)_{i_{0}j_{0}},\sigma^{2}_{g,C}\right)$ and set $\left(\bfPsi^{\prime}_{C}\right)_{i_{0}j_{0}}=\gamma$. The rest of the elements of $\bfPsi^{\prime}_{C}$ are determined from $\left(\bfPsi_{C}^{\prime}\right)^{\nu\left(G_{C}^{\prime}\right)}$ through the completion operation. We consider the candidate column precision matrix
\begin{eqnarray}\label{eq:precoladd}
 \bfK_{C}^{\prime} & = & \left(\bfQ^{*}_{C}\right)^{T}\left(\bfPsi_{C}^{\prime}\right)^{T}\bfPsi_{C}^{\prime} \bfQ^{*}_{C}.
\end{eqnarray}
\noindent We know that $\bfK^{[s]}_{C}\in \cone_{G^{[s]}_{C}}$ must satisfy $\left(\bfK_{C}^{[s]}\right)_{11}=1$. The last equality implies $\left(\bfPsi^{[s]}_{C}\right)_{11}=1/\left(\bfQ^{*}_{C}\right)_{11}$, hence $\left(\bfPsi^{\prime}_{C}\right)_{11}=1/\left(\bfQ^{*}_{C}\right)_{11}$. Therefore we have $\bfK^{\prime}_{C}\in \cone_{G^{\prime}_{C}}$ and $\left(\bfK_{C}^{\prime}\right)_{11}=1$.\\
\indent We make the transition from $\left( \bfK_{C}^{[s]},G_{C}^{[s]}\right)$ to $\left( \bfK_{C}^{\prime},G_{C}^{\prime}\right)$ with probability $\min \left\{ R_{C}^{+},1\right\}$ where
\begin{eqnarray} \label{eq:rgpluscolumn}
 R^{+}_{C} & = & \sigma_{g,C}\sqrt{2\pi}\left(\bfQ^{*}_{C}\right)_{i_{0}i_{0}}\left(\bfQ^{*}_{C}\right)_{j_{0}j_{0}}\left(\bfPsi_{C}^{[s]}\right)_{i_{0}i_{0}}\frac{I_{G^{[s]}_{C}}(\delta_{C},\bfD_{C})}{I_{G^{\prime}_{C}}(\delta_{C},\bfD_{C})} \frac{\sum\limits_{G^{\prime\prime}_{C}\in \nbd^{+}_{p_{C}}\left(G_{C}^{[s]}\right)}\left( z^{[s]} \right)^{|\nu\left( G^{\prime\prime}_{C}\right)|}}{\sum\limits_{G^{\prime\prime}_{C}\in \nbd^{-}_{p_{C}}\left(G_{C}^{\prime}\right)}\left( z^{[s]} \right)^{|\nu\left( G^{\prime\prime}_{C}\right)|}}\times \\
&&  \times \exp\left\{-\frac{1}{2}\left[\left\langle \bfK^{\prime}_{C}-\bfK^{[s]}_{C},\bfU^{*}_{C}+z^{[s]}\bfD_{C}\right\rangle -\frac{\left( \left(\Psi^{\prime}_{C}\right)_{i_{0}j_{0}}- \left(\Psi^{[s]}_{C} \right)_{i_{0}j_{0}}\right)^{2}}{\sigma_{g,C}^{2}}\right]\right\}. \nonumber
\end{eqnarray}
Next we assume that $G_{C}^{\prime}$ is obtained by deleting the edge $(i_{0},j_{0})$ from $G_{C}^{[s]}$.  We define an upper triangular matrix $\bfPsi^{\prime}_{C}$ such that $\left(\bfPsi^{\prime}_{C}\right)_{ij}=\left(\bfPsi^{[s]}_{C}\right)_{ij}$ for $(i,j)\in \nu\left(G^{\prime}_{C}\right)$. The candidate $\bfK_{C}^{\prime}$ is obtained from $\bfPsi^{\prime}_{C}$ as in equation (\ref{eq:precoladd}). We make the transition from $\left( \bfK_{C}^{[s]},G_{C}^{[s]}\right)$ to $\left( \bfK_{C}^{\prime},G_{C}^{\prime}\right)$ with probability $\min \left\{ R_{C}^{-},1\right\}$ where
\begin{eqnarray} \label{eq:rgpluscolumn}
 R^{-}_{C} & = & \left\{ \sigma_{g,C}\sqrt{2\pi}\left(\bfQ^{*}_{C}\right)_{i_{0}i_{0}}\left(\bfQ^{*}_{C}\right)_{j_{0}j_{0}}\left(\bfPsi_{C}^{[s]}\right)_{i_{0}i_{0}}\right\}^{-1}\frac{I_{G^{[s]}_{C}}(\delta_{C},\bfD_{C})}{I_{G^{\prime}_{C}}(\delta_{C},\bfD_{C})} \frac{\sum\limits_{G^{\prime\prime}_{C}\in \nbd^{-}_{p_{C}}\left(G_{C}^{[s]}\right)}\left( z^{[s]} \right)^{|\nu\left( G^{\prime\prime}_{C}\right)|}}{\sum\limits_{G^{\prime\prime}_{C}\in \nbd^{+}_{p_{C}}\left(G_{C}^{\prime}\right)}\left( z^{[s]} \right)^{|\nu\left( G^{\prime\prime}_{C}\right)|}}\times \\
&&  \times \exp\left\{-\frac{1}{2}\left[\left\langle \bfK^{\prime}_{C}-\bfK^{[s]}_{C},\bfU^{*}_{C}+z^{[s]}\bfD_{C}\right\rangle +\frac{\left( \left(\Psi^{\prime}_{C}\right)_{i_{0}j_{0}}- \left(\Psi^{[s]}_{C} \right)_{i_{0}j_{0}}\right)^{2}}{\sigma_{g,C}^{2}}\right]\right\}. \nonumber
\end{eqnarray}
We denote the updated column precision matrix and graph by $\left( \bfK_{C}^{[s+1/2]},G_{C}^{[s+1]}\right)$.\\
$ $\\
{\bf Step 4: Resample the column precision matrix}. We denote $n^{*}_{C}=np_{R}$ and $\bfU^{*}_{C}=\sum\limits_{j=1}^{n}\left( \bfx^{(j)}\right)^{T}\bfK_{R}^{[s+1]}\bfx^{(j)}$. We determine the Cholesky decomposition $\left(\bfU^{*}_{C}+z^{[s]}\bfD_{C}\right)^{-1}=\left( \bfQ^{*}_{C}\right)^{T}\bfQ^{*}_{C}$ after permuting the row and columns of $\bfU^{*}_{C}+z^{[s]}\bfD_{C}$ according to a random ordering in $\Upsilon_{p_{C}}^{(1,1)}$. The conditional distribution of $\bfK_{C}\in \cone_{G_{C}^{[s+1]}}$ with $\left(\bfK_{C}\right)_{11}=1$ is G-Wishart $\GWis_{G_{C}^{[s+1]}}\left( n^{*}_{C}+\delta_{C}, \bfU^{*}_{C} + z^{[s]}\bfD_{C}\right)$. We make the transition from $\bfK_{C}^{[s+1/2]}$ to $\bfK_{C}^{[s+1]}\in \cone_{G_{C}^{[s+1]}}$ using Metropolis-Hastings updates from Section \ref{sec:samplegwishart}. Given the notations we used in that section, we take $p=p_{C}$,  $\delta = n^{*}_{C}+\delta_{C}$, $\bfD = \bfU^{*}_{C}  + z^{[s]}\bfD_{C}$, $\bfQ = \bfQ^{*}_{C}$ and $\sigma_{m}=\sigma_{m,C}$. The constraint $\left(\bfK_{C}\right)_{11}=1$ is accommodated as described at the end of Section \ref{sec:samplegwishart}.\\
$ $\\
{\bf Step 5: Resample the auxiliary variable}. The conditional distribution of $z>0$ is
\begin{eqnarray}\label{eq:gammaz}
 \mbox{Gamma}\left( \frac{p_{C}(\delta_{C}-2)}{2}+|\nu\left( G^{[s+1]}_{C}\right)|,\frac{1}{2}\mbox{tr}\left( \bfK^{[s+1]}_{C}\bfD_{C}\right)\right).
\end{eqnarray}
Here $\mbox{Gamma}(\alpha,\beta)$ has density $f(x|\alpha,\beta) \propto \beta^{\alpha}x^{\alpha-1}\exp(-\beta x)$. We sample from (\ref{eq:gammaz}) to obtain $z^{[s+1]}$. 

\section{Sparse models for lattice data}\label{sec:spatialglm}

Conditional autoregressive (CAR) models \citep{Be74,Ma88} are routinely used in spatial statistics to model lattice data.  In the case of a single observed outcome in each region, the data is associated with a vector $\bfX^{T}=\left(X_1,\ldots,X_{p_{R}}\right)^{T}$ where $X_i$ corresponds to region $i$.  The zero-centered CAR model is implicitly defined by the set of full conditional distributions
\begin{align*}
X_i | \{ X_{i'} : i' \ne i \} & \sim \normal \left(  \sum_{i' \ne i} b_{ii'} X_{i'} , \lambda_i^2 \right), & i&=1,\ldots,p_{R}.
\end{align*}
Therefore, CAR models are just two-dimensional Gaussian Markov random fields. According to Brook's (\citeyear{Br64}) theorem, this set of full-conditional distributions implies that the joint distribution for $\bfX$ satisfies
\begin{align*}
p(\bfX | \bflambda) &\propto \exp \left\{
-\frac{1}{2} \bfX^{T} \bfLambda^{-1}(\mathbf{I}-\bfB) \bfX
\right\},
\end{align*}
\noindent where $\bfLambda = \diag\{ \lambda_1^2, \ldots, \lambda_{p_{R}}^2 \}$ and $\bfB$ is a $p_{R} \times p_{R}$ matrix such that $\bfB = (b_{ii'})$ and $b_{ii} = 0$.  In order for  $\bfLambda^{-1}(\mathbf{I}-\bfB)$ to be a symmetric matrix we require that $b_{ii'} \lambda_{i'}^2 =  b_{i'i} \lambda_i^2$ for $i \ne i'$; therefore the matrix $\bfB$ and vector $\bflambda$ must be carefully chosen.  A popular approach is to begin by constructing a symmetric proximity matrix $\bfW = (w_{ii'})$, and then set $b_{ii'} = w_{ii'}/w_{i+}$ and $\lambda_i^2 = \tau^2/w_{i+}$ where $w_{i+} = \sum_{i'} w_{ii'}$ and $\tau^{2}>0$.  In that case, $\bfLambda^{-1}(\mathbf{I}-\bfB) = \tau^{-2} (\bfE_{\bfW} - \bfW)$, where $\bfE_{\bfW} = \diag\{ w_{1 +},\ldots,w_{p_{R}+} \}$.  The proximity matrix $\bfW$ is often constructed by first specifying a neighborhood structure for the geographical areas under study; for example, when modeling state or county level data it is often assumed that two geographical units are neighbors if they share a common border.  Given that neighborhood structure, the proximity matrix is specified as
\begin{align}\label{eq:neigh}
w_{ii'} = \begin{cases}
1, & \mbox{ if }i' \in \partial i, \\
0, & \mbox{otherwise,}
\end{cases}
\end{align}
where $\partial i$ denotes the set of neighbors of $i$.

Specifying the joint precision matrix for $\bfX$ using the proximity matrix derived from the neighborhood structure is very natural; essentially, it implies that observations collected on regions that are not neighbors are conditionally independent from each other given the rest.  However, note that the specification in \eqref{eq:neigh} implies that $(\bfE_{\bfW} - \bfW) \mathbf{1}_{p_{R}} = \mathbf{0}$ and therefore the joint distribution on $\bfX$ is improper.  Proper CAR models \cite{Cr73,SuTsKiHe00,GeVo03} can be obtained by including a spatial autocorrelation parameter $\rho$, so that 
\begin{align} \label{eq:icar}
X_i | \{ X_{i'} : i' \ne i \} & \sim \normal \left( \rho \sum_{i' \ne i} \frac{w_{ii'}}{w_{i+}} X_{i'} , \frac{\tau^2}{w_{i+}} \right), 
\end{align}
The joint distribution on $\bfX$ is then multivariate normal $\normal_{p_{R}}\left( \mathbf{0},\bfD_{W}^{-1}\right)$ where $\bfD_{W}=\tau^{2}\left(\bfE_{\bfW}-\rho\bfW\right)^{-1}$. This distribution is proper as long as $\rho$ is between the reciprocals of the minimum and maximum eigenvalues for $\bfW$.  In particular, note that taking $\rho = 0$ leads to independent random effects.

In the spirit of \cite{BeKa95}, an alternative but related approach to the construction of models for lattice data is to let $\bfX \sim \normal_{p_{R}}( \mathbf{0} , \bfK^{-1} )$ and assign $\bfK$ a G-Wishart prior $\GWis_{G_{W}}\left(\delta,(\delta-2)\bfD_{W}\right)$ where the graph $G_{W}$ is implied by the neighborhood matrix $\bfW$ defined in \eqref{eq:neigh}. Following \cite{lenkoski-dobra-2010}, the mode of $\GWis_{G_{W}}\left(\delta,(\delta-2)\bfD_{W}\right)$ is the unique positive definite matrix $\bfK$ that satisfies the relations 
\begin{eqnarray}\label{eq:system}
 \bfK^{-1}_{ii^{\prime}} = \left(\bfD_{W}\right)_{ii^{\prime}}, \mbox{ if }i' \in \partial i, \mbox{ and } \bfK_{ii^{\prime}} = 0, \mbox{ if }i' \notin \partial i.
\end{eqnarray}
The matrix $\bfD_{W}^{-1}$ verifies the system (\ref{eq:system}), hence it is the mode of $\GWis_{G_{W}}\left(\delta,(\delta-2)\bfD_{W}\right)$. As such, the mode of the prior for $\bfK$ induces the same prior specification for $\bfX$ as (\ref{eq:icar}).\\
\indent  It is easy to see that, conditional on $\bfK \in \cone_{G_{W}}$, we have
\begin{eqnarray} \label{eq:icark}
X_i | \{ X_{i'} : i' \ne i \} \sim \normal \left( 
- \sum_{i' \in \partial i} \frac{K_{ii'}}{K_{ii}} X_{i'}, \frac{1}{K_{ii}}
\right).
\end{eqnarray}
Hence, by modeling $\bfX$ using a Gaussian graphical model and restricting the precision matrix $\bfK$ to belong to the cone $\cone_{G_{W}}$, we are inducing a mixture of CAR priors on $\bfX$ where the priors on
$$
b_{ii'} =\begin{cases}
 - K_{ii'}/K_{ii}, & \mbox{ if }i' \in \partial i, \\
 0, & \mbox{ otherwise, }
 \end{cases}
$$
and $\lambda_i^2 = 1/K_{ii}$ are induced by the G-Wishart prior $\GWis_{G_{W}}\left(\delta,(\delta-2)\bfD_{W}\right)$.

The specification of CAR models through G-Wishart priors solves the impropriety problem and preserves the computational advantages derived from standard CAR specifications while providing greater flexibility.  Indeed, the prior is trivially proper because the matrix $\bfK \in \cone_{G_{W}}$ is invertible by construction.  The computational advantages are preserved because the full conditional distributions for each $X_i$ can be easily computed for any matrix $\bfK$ without the need to perform matrix inversion, and they depend only on a small subset of neighbors $\{ X_j : j \in \partial i \}$.  Additional flexibility is provided because the weights $b_{ii'}$ for $i' \in \partial i$ and smoothing parameters $\lambda_i^2$ are being estimated from the data rather than being assumed fixed, allowing for adaptive spatial smoothing.  
Indeed, our approach provides what can be considered as a nonparametric alternative to the parametric estimates of the proximity matrix proposed by \cite{CrCh89}.  

A similar approach can be used to construct proper multivariate conditional autoregressive (MCAR) models \citep{Ma88,GeVo03}.  In this case, we are interested in modeling a $p_R \times p_C$  matrix $\bfX = (X_{ij})$ where $X_{ij}$ denotes the value of the $j$-th outcome in region $i$.  We let $\bfX$ follow a matrix-variate normal distribution with row precision matrix $\bfK_R$ capturing the spatial structure in the data (which, as in univariate CAR models, is restricted to the cone $\cone_{G_{W}}$ defined by the neighborhood matrix $\bfW$), and column precision matrix $\bfK_C$, which controls the multivariate dependencies across outcomes.  It can be easily shown that the row vector $\bfX_{i\textasteriskcentered}$ of $\bfX$ depends only on the row vectors associated with those regions that are neighbors with $i$ --- see also equation (\ref{eq:matvarci}):
\begin{align*}
\bfX_{i\textasteriskcentered}^{T} | \left\{ \bfX_{i'\textasteriskcentered}^{T} : i' \ne i \right\} \sim \normal_{p_{C}} \left( 
- \sum_{i' \in \partial i} \frac{ (\bfK_R)_{ii'}}{(\bfK_R)_{ii}} \bfX_{i'\textasteriskcentered}^{T}, \frac{1}{(\bfK_R)_{ii}} \bfK_C^{-1}
\right).
\end{align*}
\noindent This formulation for spatial models can also be used as part of more complex hierarchical models.  Indeed, CAR and MCAR models are most often used as a prior for the random effects of a generalized linear model (GLM) to account for residual spatial structure not accounted for by covariates.  When no covariates are available, the model can be interpreted as a spatial smoother where the spatial covariance matrix controls the level of spatial smoothing in the underlying (latent) surface.  Similarly, MCAR models can be used to construct sparse multivariate spatial GLMs.

As an example, consider the $p_{R}\times p_{C}$ matrix  $\bfY = \left(Y_{ij}\right)$  of discrete or continuous outcomes, and let $Y_{ij} \sim h_{j}( \cdot | \eta_{ij} )$ where $h_j$ is a probability mass or probability density function that belongs to the exponential family with location parameter $\eta_{ij}$.  The spatial GLM is then defined through the linear predictor
$$
g^{-1}(\eta_{ij})  = \mu_j + X_{ij} + \bfZ_{ij} \bfbeta_{j},
$$
where $g$ is the link function, $\mu_j$ is an outcome-specific intercept, $X_{ij}$ is a zero-centered spatial random effect associated with location $i$, $\bfZ_{ij}$ is a matrix of observed covariates for outcome $j$ at location $i$, and $\bfbeta_{j}$ is the vector of fixed effects associated with outcome $j$.  We could further take $y_{ij} \sim \Poi(\eta_{ij})$, $g^{-1} (\cdot) = \log(\cdot)$ and assign a sparse matrix-variate normal distribution for $\bfX$.  This choice leads to a sparse multivariate spatial log-linear model for count data, which is often used for disease mapping in epidemiology (see Section \ref{se:diseasemap}).


Due to the structure associated with our centering matrix $\bfD_{R}$, and the presence of the expanded scaling parameter $z$, prior elicitation for this class of spatial models is relatively straightforward.  Indeed, elicitation of the matrix $\bfD_{R}$ requires only the elicitation of the neighborhood matrix $\bfW$, along with reasonable values for the spatial autocorrelation parameter $\rho$ and scale $\tau^2$.  In particular, in the applications we discuss below we assume that, {\it a priori}, there is a strong degree of positive spatial association, and set $\rho = 0.99$.  Also, and as is customary in the literature on spatial models for lattice data, $\tau^2$ is selected to roughly reflect the scale of data being modeled. In the case of MCAR models, it is common to assume that the prior value for the conditional covariance between variables is zero, which leads to choosing a diagonal $\bfD_C$.  When the scale of all the outcome variables is the similar (as in the applications below), it is then sensible to pick $\bfD_C = \mathbf{I}_{p_C}$, as the actual scale is irrelevant.

\section{Illustrations}\label{sec:illust}

\subsection{Simulation study}\label{sec:sim_study}

\indent We consider a simulation study involving matrix-variate normal graphical models where $p_{R} = 5$ and $p_{C} = 10$.  We sample $100$ observations of matrices with row graph $G_R$ defined by the edges $\{(1,i): 1 < i \leq 5\} \cup \{(2,3),(3,4),(4,5),(2,5)\}$ and column graph $C_{10}$.  Here $C_{10}$ denotes the 10-dimensional cycle graph with edges $\{(i,i + 1): 1 \leq i \leq 9\} \cup \{(1,10)\}$.  We note that both the row and column graphs are non-decomposable, with the row graph being relatively dense, while column graph is relatively sparse.  In this case $\bfK_R \in \cone_{G_R}$ is set so that $(\bfK_{R})_{ii} = 1$, $(\bfK_{R})_{i, j} = 0.4$ for $(i,j)\in G_R$. $\bfK_C \in \cone_{C_{10}}$ is defined in a similar manner.  To generate the $100$ samples from this matrix-variate distribution, we sample $\mbox{vec}(\bfX^{T})\sim \normal_{p_{R}p_{C}}\left(\bf0,\left[\bfK_{R}\otimes \bfK_{C}\right]^{-1}\right)$ and rearrange this vector to obtain the appropriate matrices.\\
\indent We ran the matrix-variate GGM search described in Section \ref{sec:matrixggm} for 50,000 iterations after a burn in of 5,000 iterations.  We set $\sigma_{m,R}$, $\sigma_{m,C}$, $\sigma_{g,R}$ and $\sigma_{g,C}$ all to the value $.5$, which achieved an average rejection rate on updates of elements for both $\bfK_C$ and $\bfK_R$ of about 30 percent.  We repeated the study $100$ times to reduce sampling variability.  Tables \ref{tab:sim_row} and \ref{tab:sim_col} show the resulting estimates for the row and column GGMs respectively.  We see that the matrix-variate GGM search outlined above performs extremely well at recovering the structure of both the column and row graphs as well as the values of both $\bfK_C$ and $\bfK_R$.  Since both row and column GGMs are defined by non-decomposable graphs, we have illustrated that our methodology is capable of searching the entire graph space of matrix-variate GGMs.\\
\begin{table}\caption{Results of simulation study in estimates for the row graph and $\bfK_{R}$.  In the matrix below, the values along the diagonal and in the upper triangle correspond to the average of the estimates of entries in $\bfK$ across the $100$ repetitions, while the values below the diagonal correspond to the average of the edge probabilities across the $100$ iterations.}\label{tab:sim_row}
\vspace{.1cm}
\begin{center}
\begin{tabular}{c |c c c c c}
\hline\hline
0 & 1 & 2 & 3 & 4 & 5\\
\hline
1 & 1.107 & 0.442 & 0.438 & 0.436 & 0.44\\
2 & 1 & 1.105 & 0.439 & 0 & 0.441\\
3 & 1 & 1 & 1.1 & 0.437 & 0\\
4 & 1 & 0.026 & 1 & 1.086 & 0.438\\
5 & 1 & 1 & 0.04 & 1 & 1.096\\
\hline\hline
\end{tabular}
\end{center}
\end{table}

\begin{table}\caption{Results of simulation study for estimates of the column graph and $\bfK_{C}$.  In the matrix below, the values along the diagonal and above it to the average of the estimates of entries in $\bfK_C$ across the $100$ repetitions, while the values below the diagonal correspond to the average of the edge probabilities across the $100$ iterations.}\label{tab:sim_col}
\begin{center}
\begin{tabular}{c | c c c c c c c c c c}
\hline\hline
 & 1 & 2 & 3 & 4 & 5 & 6 & 7 & 8 & 9 & 10\\
\hline
1 & 1 & 0.395 & -0.001 & 0.001 & 0 & -0.001 & 0 & 0.001 & -0.001 & 0.396\\
2 & 1 & 0.941 & 0.364 & 0 & 0.001 & 0 & 0 & 0.001 & 0.002 & 0.001\\
3 & 0.024 & 1 & 0.916 & 0.365 & 0 & 0.001 & -0.001 & 0 & 0 & 0\\
4 & 0.046 & 0.032 & 1 & 0.932 & 0.37 & -0.001 & 0 & 0 & 0.001 & -0.001\\
5 & 0.022 & 0.046 & 0.033 & 1 & 0.936 & 0.371 & -0.001 & 0 & -0.001 & 0.001\\
6 & 0.04 & 0.038 & 0.036 & 0.034 & 1 & 0.927 & 0.362 & -0.001 & 0 & -0.001\\
7 & 0.037 & 0.03 & 0.052 & 0.025 & 0.042 & 1 & 0.928 & 0.366 & 0 & 0\\
8 & 0.047 & 0.046 & 0.024 & 0.053 & 0.032 & 0.049 & 1 & 0.935 & 0.377 & 0\\
9 & 0.072 & 0.082 & 0.039 & 0.029 & 0.036 & 0.068 & 0.039 & 1 & 0.93 & 0.364\\
10 & 1 & 0.057 & 0.037 & 0.03 & 0.047 & 0.044 & 0.035 & 0.066 & 1 & 0.946\\
\hline\hline
\end{tabular}
\end{center}
\end{table}


\subsection{Modeling statewide SAT scores}\label{se:sat}

\begin{figure}[!h]
\begin{center}
\caption{SAT scores in the 48 contiguous states.  Left panel shows verbal scores, while the right panel shows mathematics scores.}\label{fi:SATmaps}
\subfigure[Verbal]{\includegraphics[width=3.1in,angle=0]{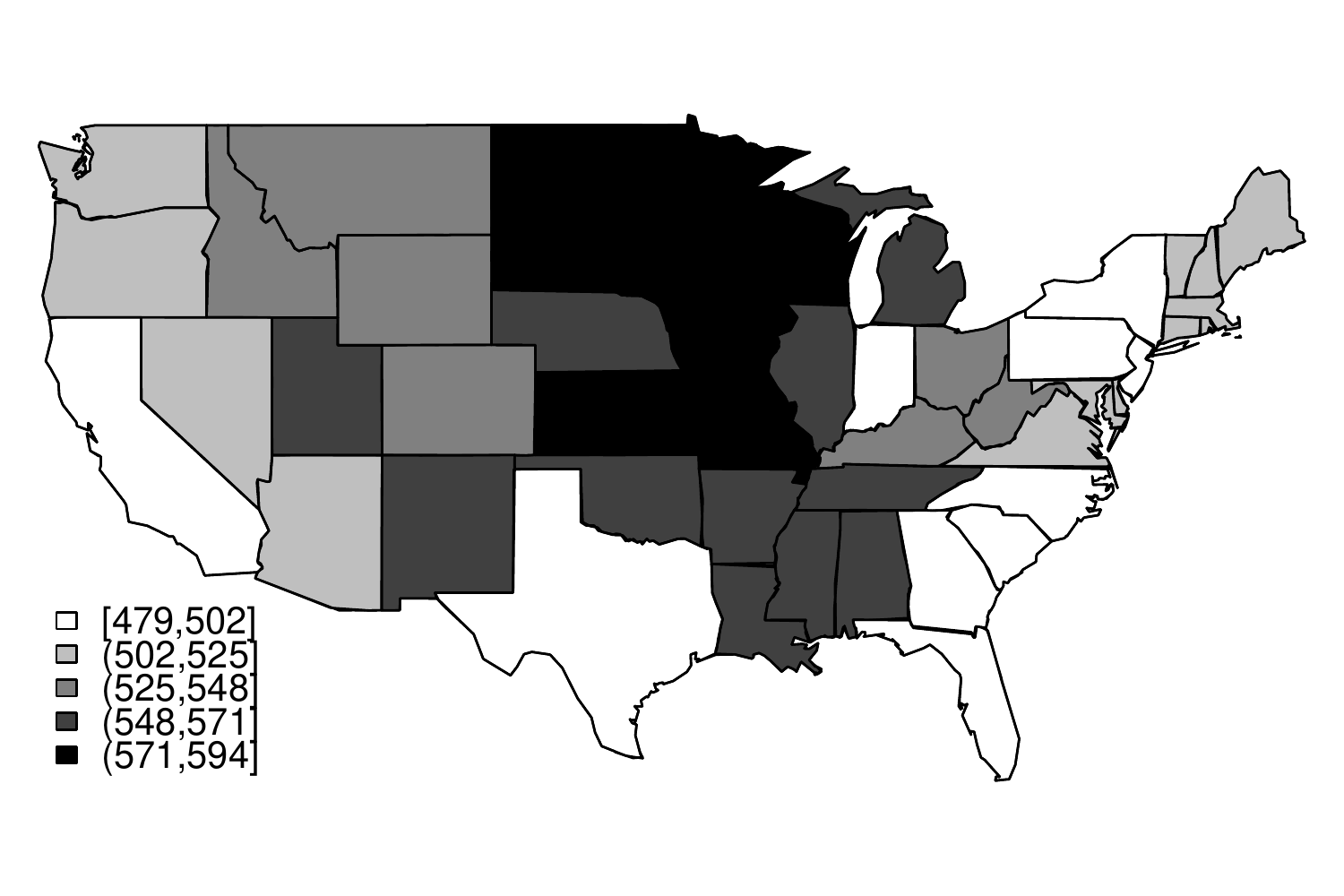}\label{fi:verbalraw}} 
\subfigure[Mathematics]{\includegraphics[width=3.1in,angle=0]{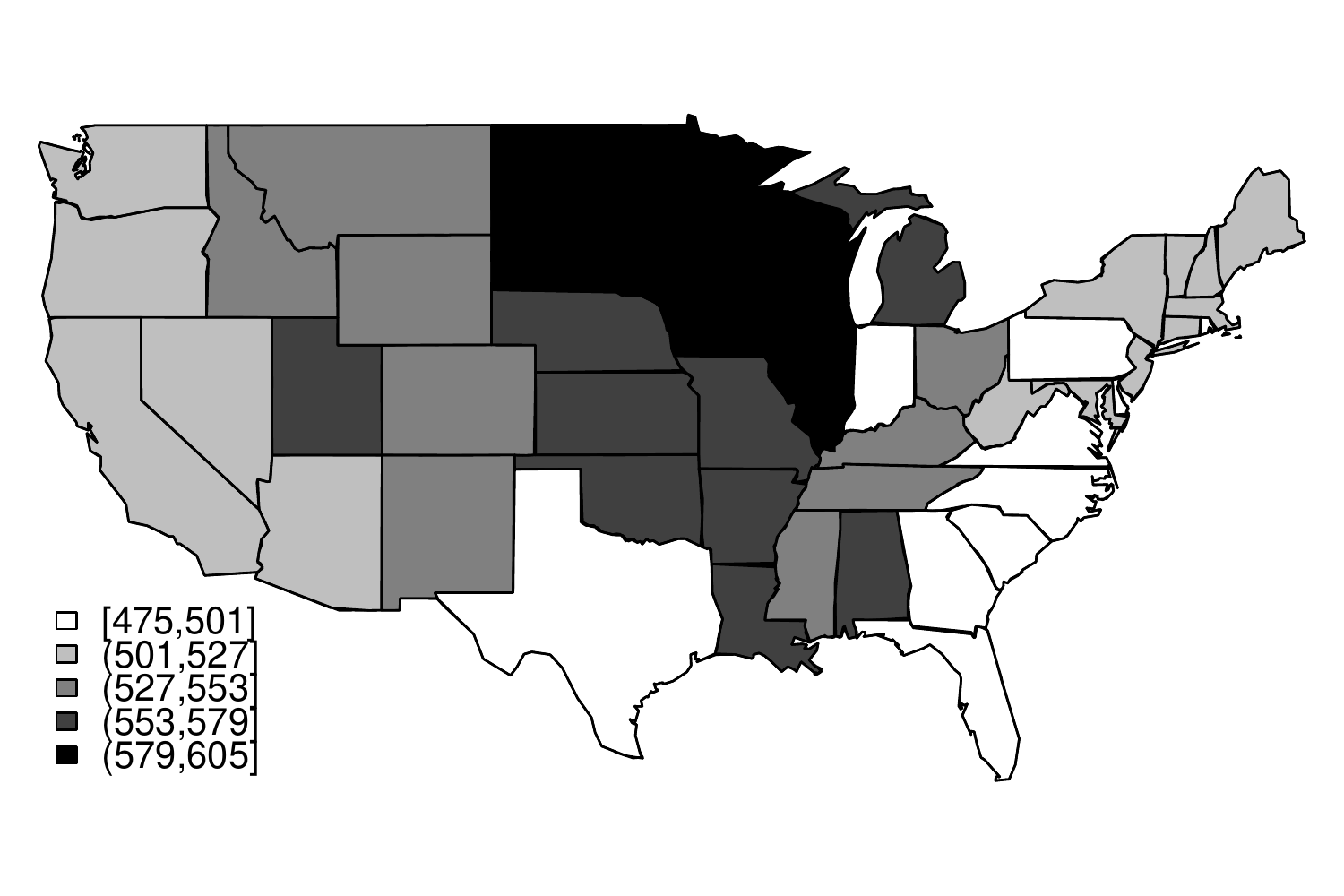}\label{fi:mathraw}}
\end{center}
\end{figure}

This section develops a model for the scores on the SAT college entrance exam for the year 1999.  The data consists of statewide average verbal and mathematics SAT scores, along with the percentage of eligible students taking the test, for each of the 48 contiguous states plus the District of Columbia.  The verbal scores were originally analyzed by \cite{Wa04}, who used them to illustrate the differences among different models for lattice data. Figure \ref{fi:SATmaps} shows choropleth maps of the two scores; in both cases there is strong evidence of spatial association (note the clear clusters of midwestern, central, western and southeastern states).

\begin{figure}[!h]
\begin{center}
\caption{Scatterplots for the SAT data set.  Panel (a) shows the average verbal SAT score in the state vs. the number of eligible students that take the test, panel (b) shows the average math SAT score against the number of eligible students, and panel (c) shows the average verbal vs. the average math scores.}\label{fi:SATscatter}
\subfigure[]{\includegraphics[width=2.1in,angle=0]{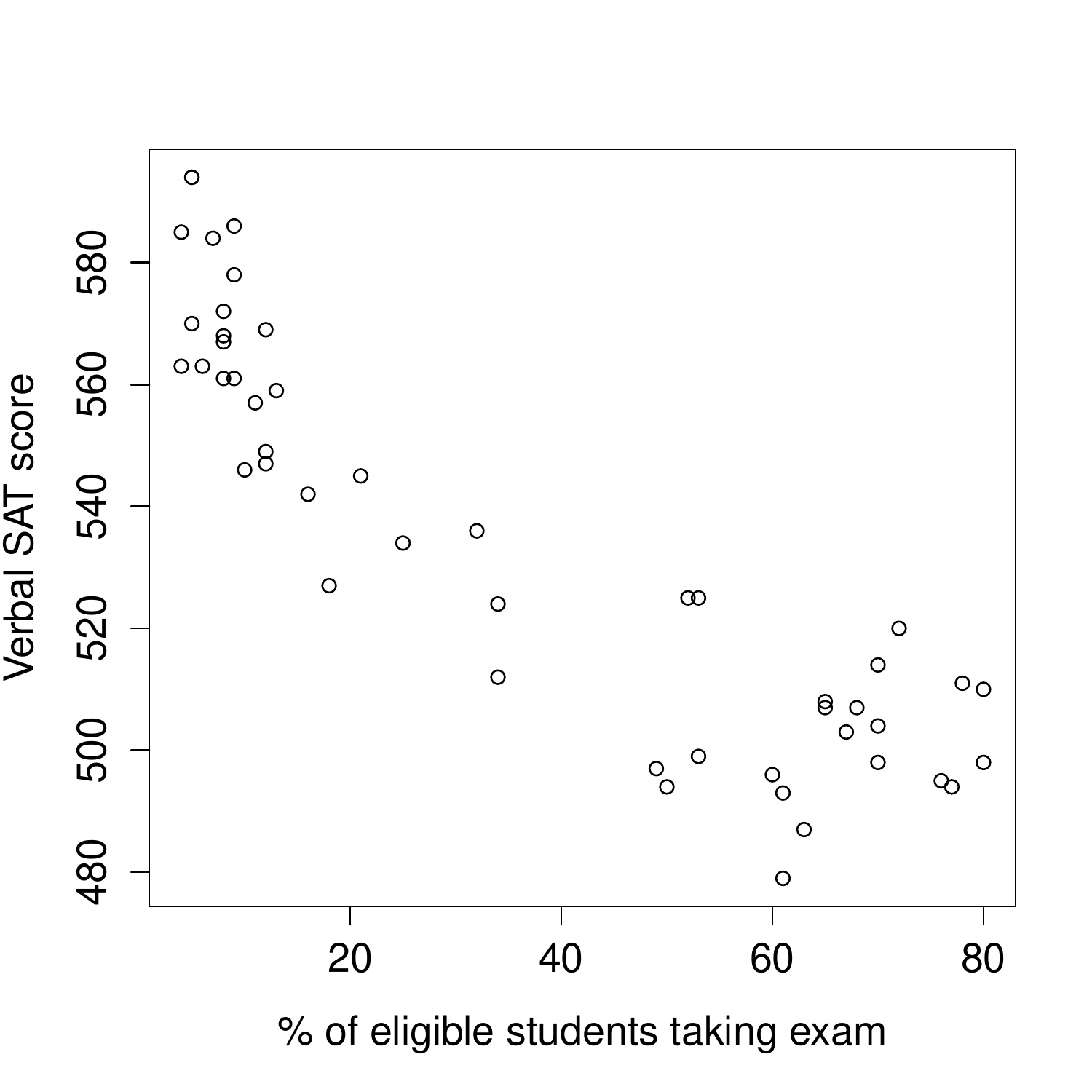}\label{fi:verbalpercent}} 
\subfigure[]{\includegraphics[width=2.1in,angle=0]{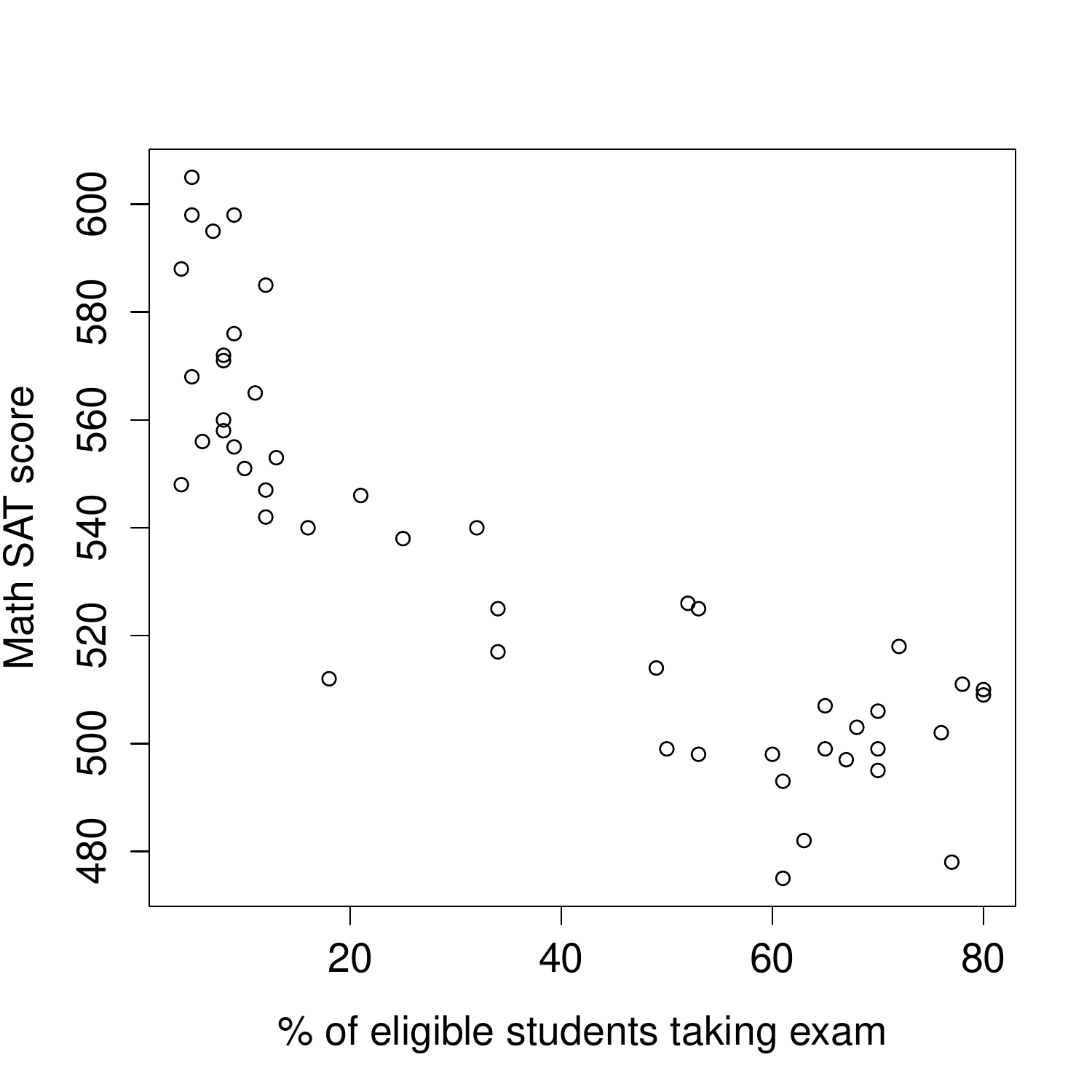}\label{fi:mathpercent}}
\subfigure[]{\includegraphics[width=2.1in,angle=0]{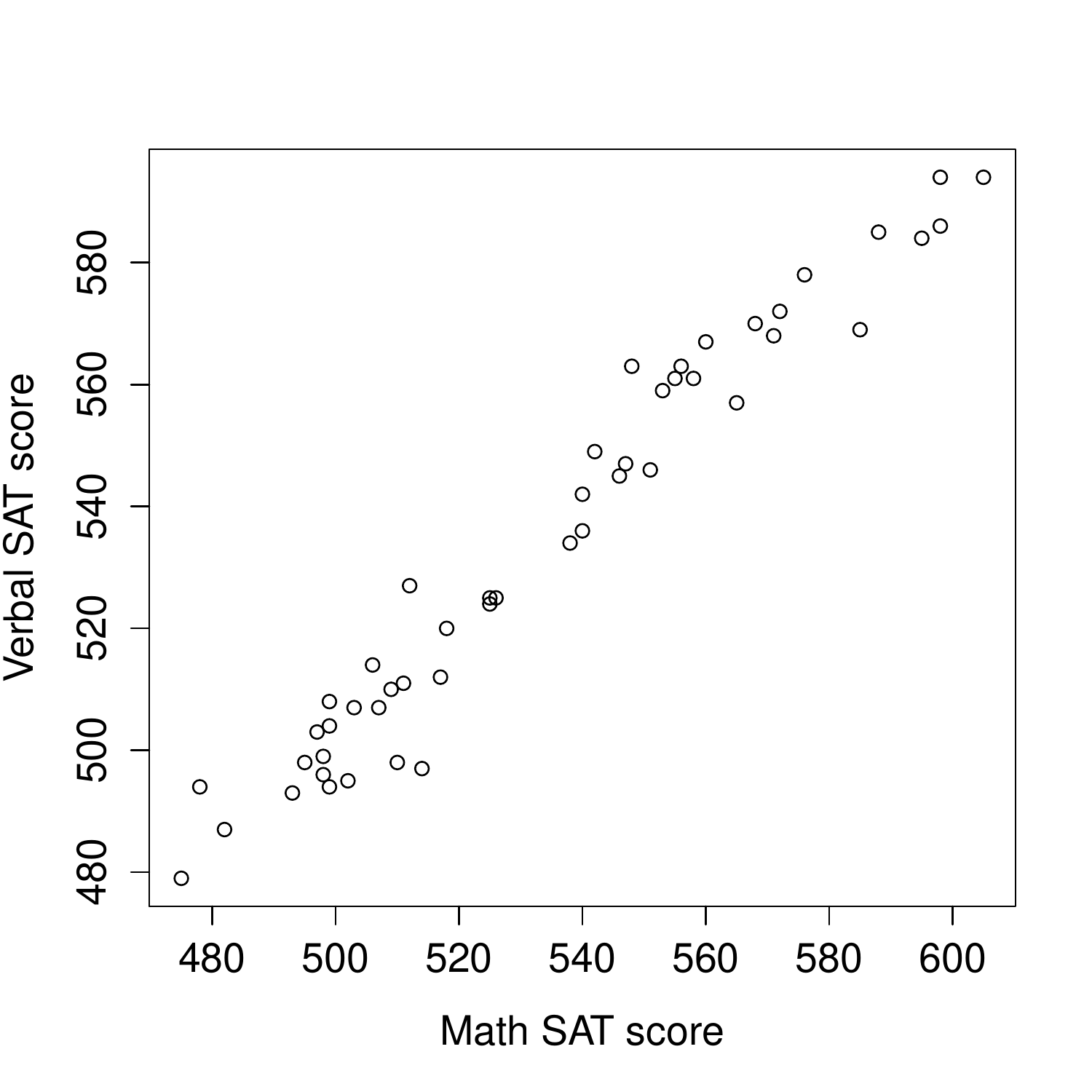}\label{fi:mathverbal}}
\end{center}
\end{figure}

In this section we develop a bivariate spatial regression where both verbal and mathematics scores are treated as response variables, while the percentage of eligible students are treated as the explanatory variable.  Figure \ref{fi:SATscatter} presents pairwise scatterplots of all three variables.  They demonstrate that both verbal and mathematics scores are positively associated with each other, and negatively associated with the percentage of eligible students who take the test.  This association between SAT scores and the percentage of eligible students could be explained by selection bias: in states where fewer students take the exam, it is typically only the best qualified who do, which in turn results in better averages than those from states who provide a wider access to testing.  Hence, our model attempts to jointly understand the behavior of the scores adjusting for this selection bias.

\begin{figure}[!h]
\begin{center}
\caption{Fitted SAT scores in the 48 contiguous states.  Left panel shows verbal scores, while the right panel shows mathematics scores. The scale is the same as in Figure~\ref{fi:SATmaps} }\label{fi:SATfittedmaps}
\subfigure[Verbal]{\includegraphics[width=3.1in,angle=0]{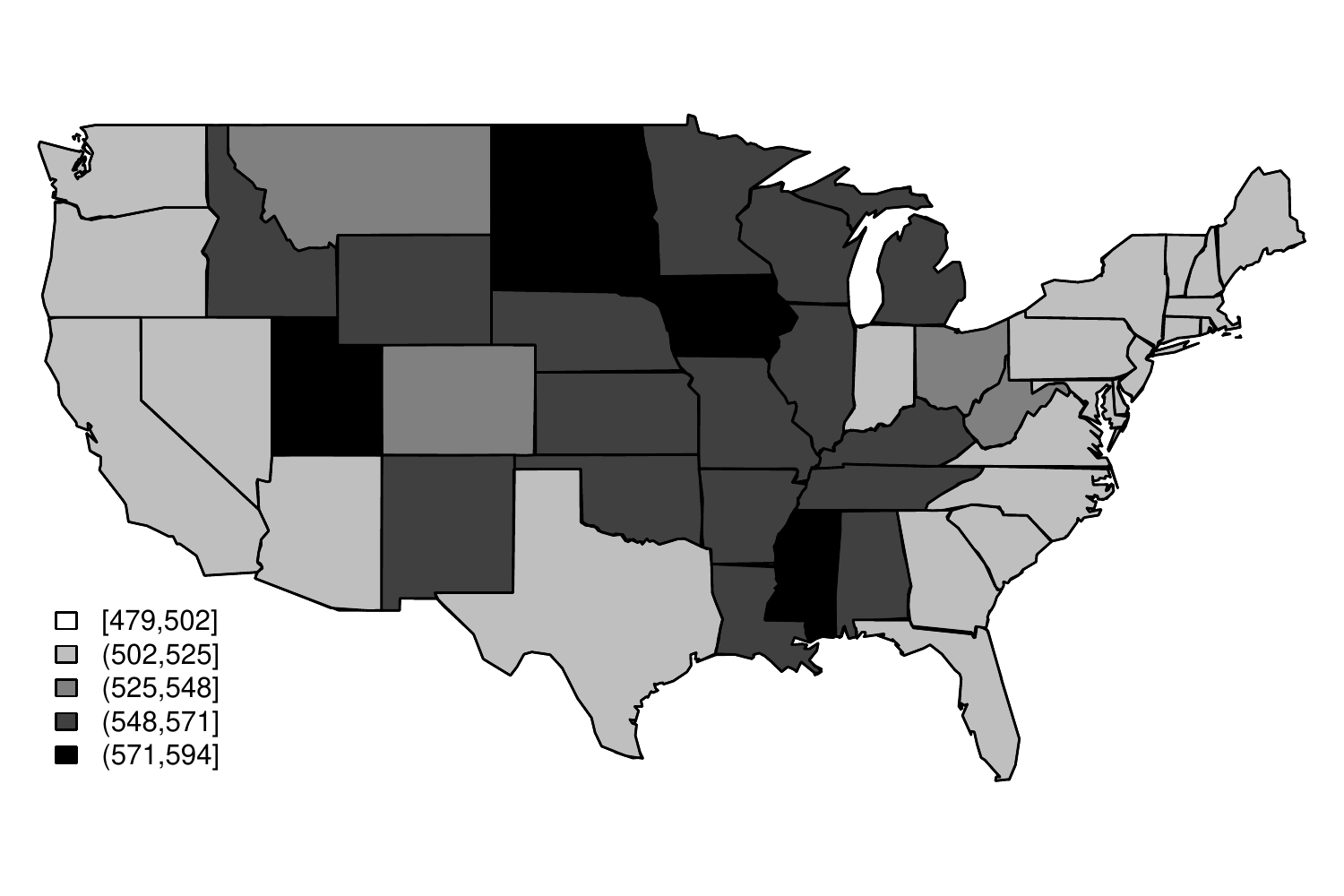}\label{fi:verbalfitted}} 
\subfigure[Mathematics]{\includegraphics[width=3.1in,angle=0]{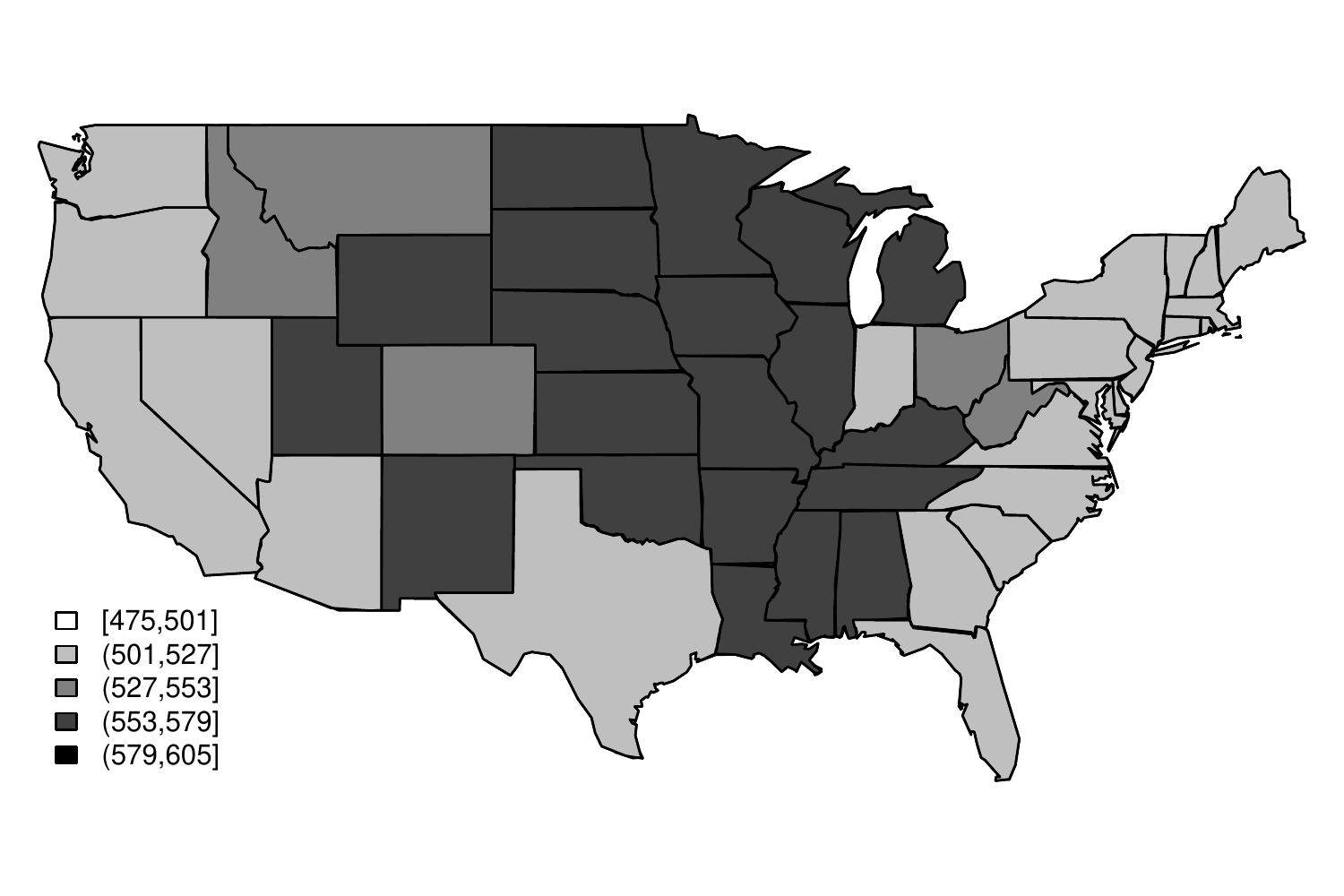}\label{fi:mathfitted}}
\end{center}
\end{figure}

\begin{figure}[!h]
\begin{center}\caption{A comparison of $\bfD_{W}^{-1}$ and the posterior mean of $\bfK_R$ in the SAT example.  For comparison, the heat maps have the same scale, showing the considerable added variability in spatial dependence imposed by the model.}\label{fi:SATheatmaps}
\subfigure[Prior]{\includegraphics[width=3.1in, angle=0]{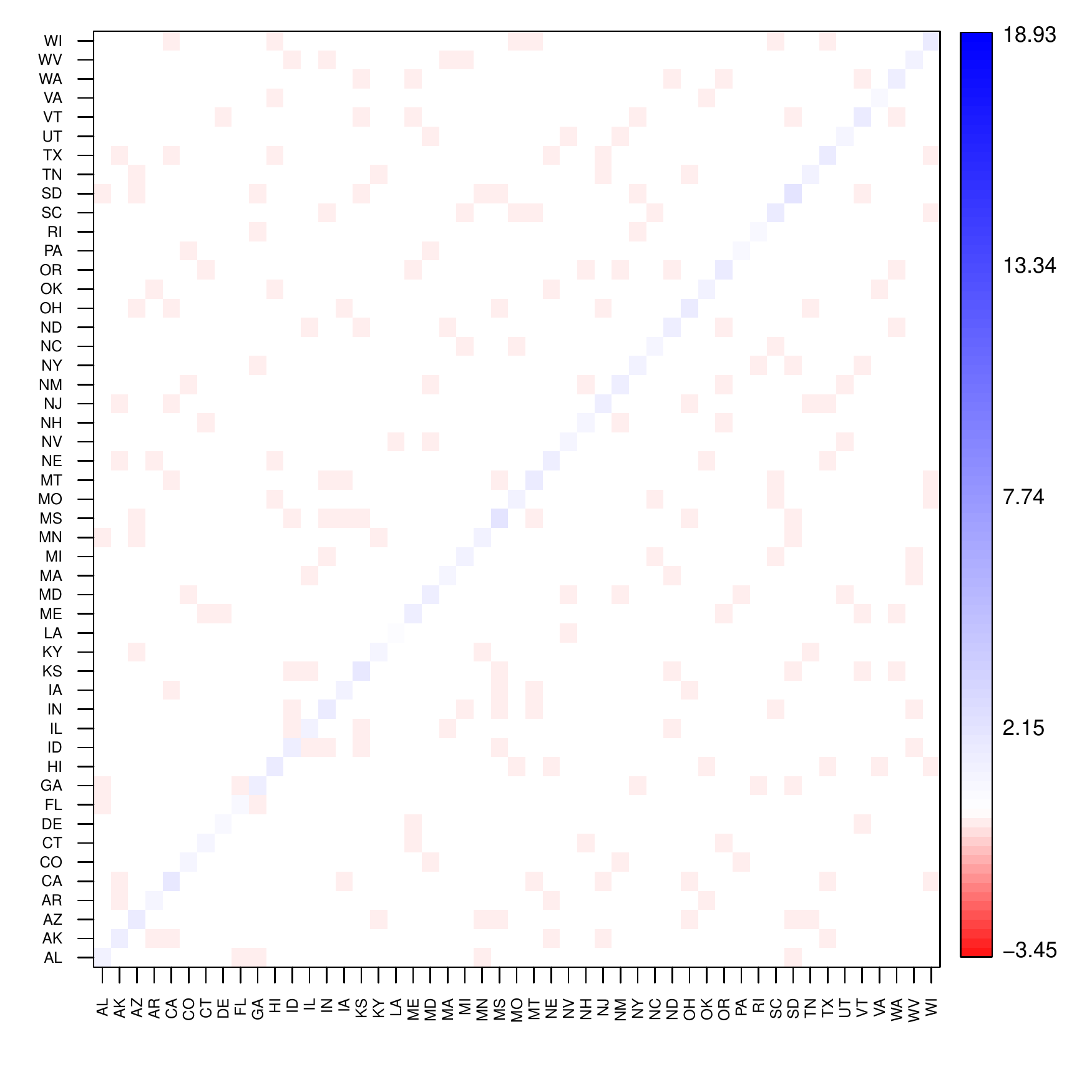}\label{fi:SATheatmapprior}}
\subfigure[Posterior]{\includegraphics[width=3.1in, angle=0]{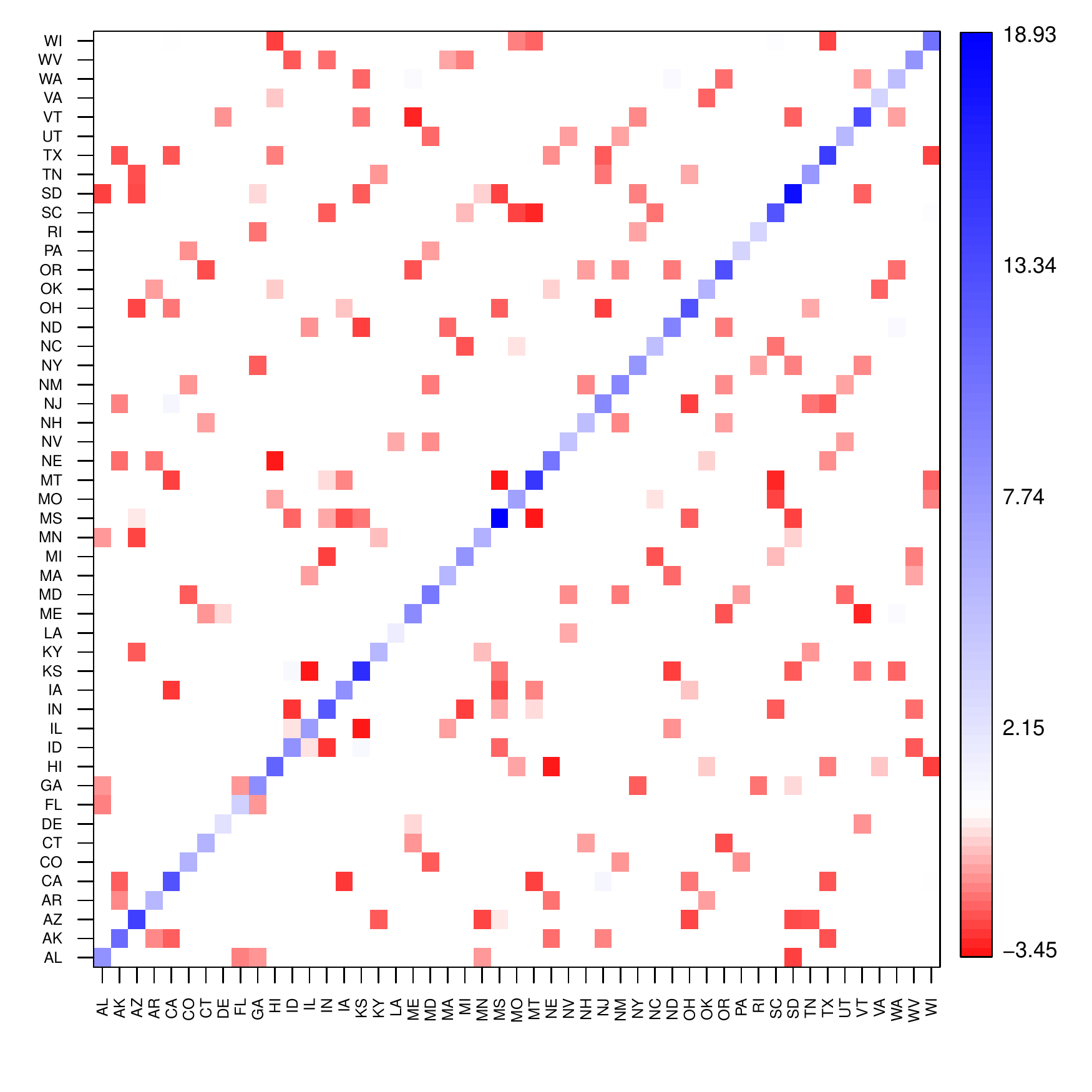}\label{fi:SATheatmapost}}
\end{center}
\end{figure}

More specifically, we let $Y_{ij}$ be the score on test $j$ ($j=1$ for verbal and $j=2$ for math) and state $i = 1,\ldots, 49$.  We use the notation from Section \ref{sec:matrixggm} and take $p_{R}=49$, $p_{C}=2$. The observed scores form a $p_{R}\times p_{C}$ matrix $\bfY=\left( Y_{ij}\right)$. Since Figure \ref{fi:SATscatter} suggests a quadratic relationship between scores and the percentage of eligible students, we let 
$$Y_{ij} = \beta_{0j} + \beta_{1j} Z_{i} + \beta_{2j}Z_{i}^2 + X_{ij},
$$
\noindent where $Z_{i}$ represents the percentage of eligible students taking the exam and $X_{ij}$ is an error term. The vectors $\bfbeta_{j}=\left( \beta_{0j},\beta_{1j},\beta_{2j}\right)^{T}$, $j=1,2$, are given independent normal priors $\normal_{3}\left( \bfb_{j},\bfOmega_{j}^{-1}\right)$ with $\bfb_{j}=(550, 0, 0)^{T}$ and $\bfOmega_{j}^{-1}=\diag\{ 225, 10, 10\}$.  The residual matrix $\bfX = (X_{ij})$ is given a matrix variate GGM
\begin{eqnarray*}
 \mbox{vec}\left(\bfX^{T}\right)|\bfK_{R},\bfK_{C} & \sim & \normal_{p_{R}p_{C}}\left( \mathbf{0} ,\left[\bfK_{R}\otimes \bfK_{C}\right]^{-1}\right),\\
 \mathbf{K}_{R}|\delta_{R},\bfD_{R} & \sim & \GWis_{G_{R}}\left(\delta_{R},\bfD_{R}\right),\\
 \left(z\mathbf{K}_{C}\right)|\delta_{C},\bfD_{C} & \sim & \GWis_{G_{C}}\left(\delta_{C},\bfD_{C}\right). 
 \end{eqnarray*}
\noindent where $z>0$  is an auxiliary variable needed to impose the identifiability constraint $\left( \bfK_{C}\right)_{11}=1$. The graph $G_R$, which captures the spatial dependencies among states, is fixed to the one induced by the neighborhood matrix $\bfW = (w_{ii'})$, where $w_{ij}=1$ if and only if states $i$ and $i'$ share a border, while we set a uniform prior on $G_C$.  The degrees of freedom for the G-Wishart priors are set as $\delta_C = \delta_R = 3$, while the centering matrices are chosen as $\bfD_C = \mathbf{I}_{2}$ and, in the spirit of CAR models, $\bfD_R = (\delta_{R}-2)w_{1+} ( \bfE_{\bfW} - 0.99\bfW)^{-1}$. It follows that
\begin{eqnarray} \label{eq:ysat}
  \mbox{vec}\left(\bfY^{T}\right)|\bfK_{R},\bfK_{C} \sim \normal_{p_{R}p_{C}}\left( \mbox{vec}\left(\left(\bfZ\bfbeta\right)^{T}\right),\left[\bfK_{R}\otimes \bfK_{C}\right]^{-1}\right),
\end{eqnarray}
where $\bfZ$ is a $p_{R}\times 3$ matrix whose $i$-th row corresponds to the vector $(1, Z_{i}, Z_{i}^2)$ and $\bfbeta$ is a $3\times 2$ matrix whose $j$-th column is $\bfbeta_{j}$.\\
\indent A MCMC algorithm for sampling from the posterior distribution of this model's parameters is described in Appendix \ref{ap:MCMCsat}.  This algorithm was run for $20000$ iterations after a burn in of $1000$ iterations.  To assess convergence, we ran 10 instances of the algorithm at these settings, each initiated at different starting points and with different initial random seeds.  Figure~\ref{fi:converge} shows the mean value, by iteration of the parameter $z$ for each chain, which display significant agreement.\\
\indent Table~\ref{tab:sat_est} shows the posterior estimates for the regression coefficients $\bfbeta_{0}$, $\bfbeta_1$ and $\bfbeta_2$, along with the values reported by \cite{Wa04} for the verbal scores using two different models for the spatial structure:  1) a conventional CAR model, 2) an isotropic exponential variogram where the centroid of each state is used to calculate distances between states (see \citealp{Wa04} for details).  Note that the posterior means obtained from our model roughly agree with those produced by the more conventional approaches to spatial modeling.  However, the Bayesian standard errors obtained from our model are consistently larger than the standard errors reported in \cite{Wa04}.  We believe that this can be attributed to the richer spatial structure in our model, with the uncertainty in the spatial structure manifesting itself as additional uncertainty in other model parameters.\\
\indent As expected, the data suggest a positive association between SAT scores. The posterior probability of an edge between verbal and math scores is 80\%, suggesting moderate evidence of association between scores, while the posterior mean for the correlation between scores (conditional on it being different from zero) is 0.35 with a 95\% credible interval of $(0.09, 0.58)$.\\
\indent Figure~\ref{fi:SATfittedmaps} shows the fitted values of math and verbal scores by state.  In comparison to Figure~\ref{fi:SATmaps}, we see considerable shrinkage of the fitted scores, particularly for the mathematics SAT. 
We note that, even after accounting for the percentage of students taking the exam, substantial spatial structure remains; generally speaking, central states (except Texas) tend to have higher SAT scores than coastal states on either side.  This is confirmed by the results in Figure~\ref{fi:SATheatmaps}, which compares the spatial interaction structure implied by the prior parameter $\bfD_R^{-1}$ to the structure implied by the posterior mean of $\bfK_R$.  Since both matrices share the same underlying conditional independence graph, both matrices have the same pattern of zeroes.  If no spatial structure were present in the data we would expect the non-zero entries to have shrunk towards zero.  However the posterior for $\bfK_R$ shows considerably larger variation in the spatial relationship structure than the prior.  Indeed, while the prior imposes negative values for the off-diagonal of $\bfD_R^{-1}$ (implying a positive correlation), the posterior of $\bfK_R$ has several positive values, the largest of which is the entry between Colorado and New Mexico, implying a negative correlation between deviations in the SAT scores of these two states. More generally, note that the posterior mean for $\bfK_R$ is very different from its prior mean, which suggest that the model is able to learn from the data about the spatial correlation structure.\\

\begin{table}\caption{Posterior Estimates for the regression parameters in the SAT data.  The columns labeled GGM-MCAR correspond to the estimates obtained from the bivariate CAR model described in Section \ref{se:sat}, while the column labeled Wall partially reproduces the results from \cite{Wa04} for the verbal SAT scores.
}\label{tab:sat_est}
\begin{center}
\begin{tabular}{lcc|cc}
\hline\hline
& \multicolumn{2}{c}{GGM-MCAR} & \multicolumn{2}{c}{Wall (Verbal)} \\
& Math & Verbal  & CAR & Variogram\\
\hline
Intercept ($\beta_{0j}$) & 580.12 & 583.26 & 584.63 & 583.64 \\
 & (17.3) & (15.91) & (4.86) & (6.49)\\
Percent Eligible ($\beta_{1j}$) & -2.147 & -2.352 & -2.48 & -2.19\\
 & (0.394) & (0.498) & (0.32) & (0.31) \\
Percent Eligible Squared ($\beta_{2j}$)& 0.0154 & 0.0171 & 0.0189 & 0.0146\\
 & (0.0055) & (0.0069) & (0.004) & (0.004) \\
\hline\hline
\end{tabular}
\end{center}
\end{table}

\subsection{Mapping cancer mortality in the U.S.}\label{se:diseasemap}

Accurate and timely counts of cancer mortality are very useful in the cancer surveillance community for purposes of efficient resource allocation and planning. Estimation of current and future cancer mortality broken down by geographic area (state) and tumor have been discussed in a number of recent articles, including \cite{TiGhJeHaWaThFe04}, \cite{GhTi07}, \cite{GhTiFeCrJe07} and \cite{GhGhTi08}.  This section considers a multivariate spatial model on state-level cancer mortality rates in the United States for 2000.  These mortality data are based on death certificates that are filed by certifying physicians and is collected and maintained by the National Center for Health Statistics (\url{http://www.cdc.gov/nchs}) as part of the National Vital Statistics System. The data is available from the Surveillance, Epidemiology and End Results (SEER) program of the National Cancer Institute (\url{http://seer.cancer.gov/seerstat}).

\begin{figure}[!h]
\begin{center}
\subfigure[Colon]{\includegraphics[width=3.1in,angle=0]{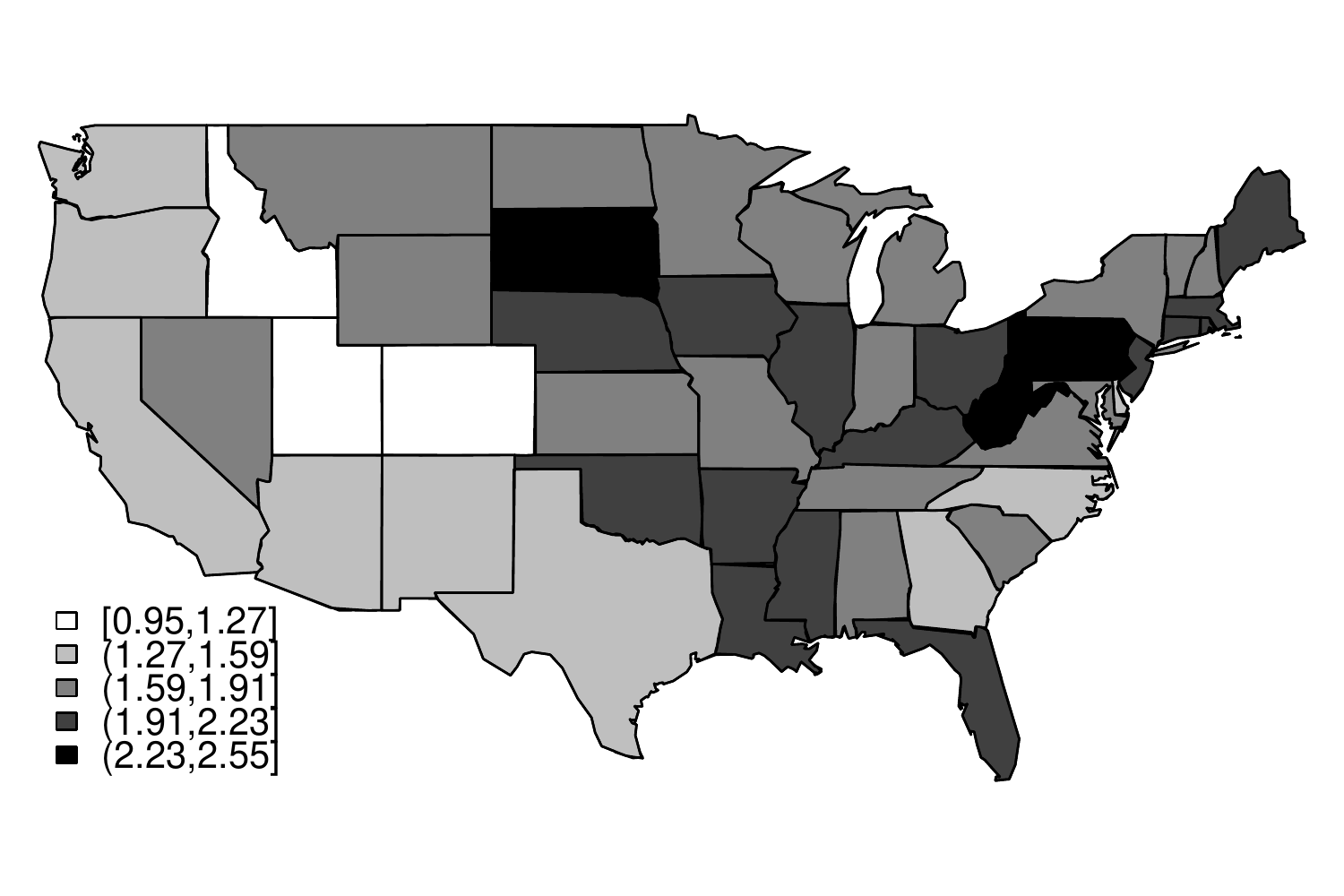}\label{fi:colonraw}} 
\subfigure[Lung]{\includegraphics[width=3.1in,angle=0]{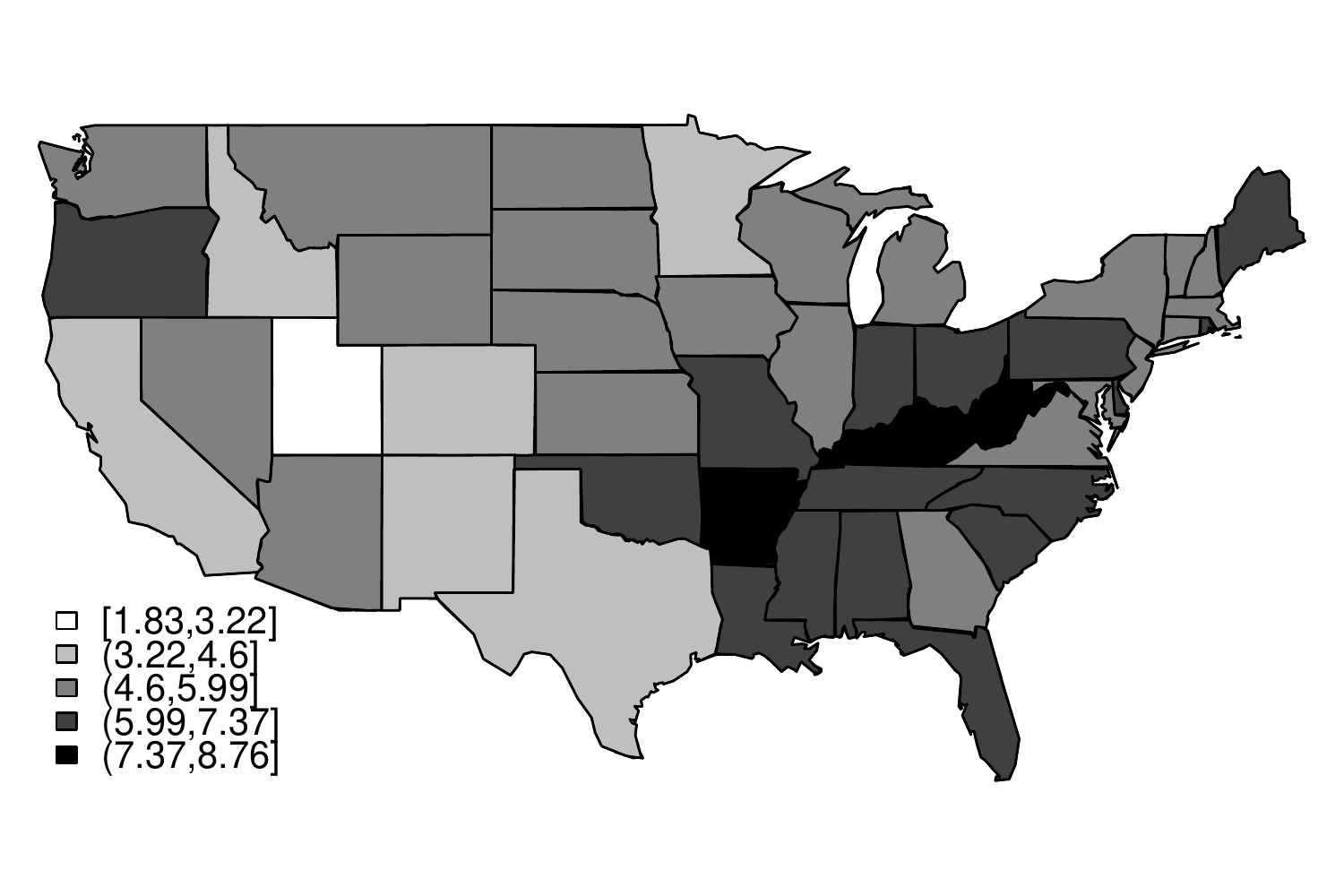}\label{fi:lungraw}}
\subfigure[Breast]{\includegraphics[width=3.1in,angle=0]{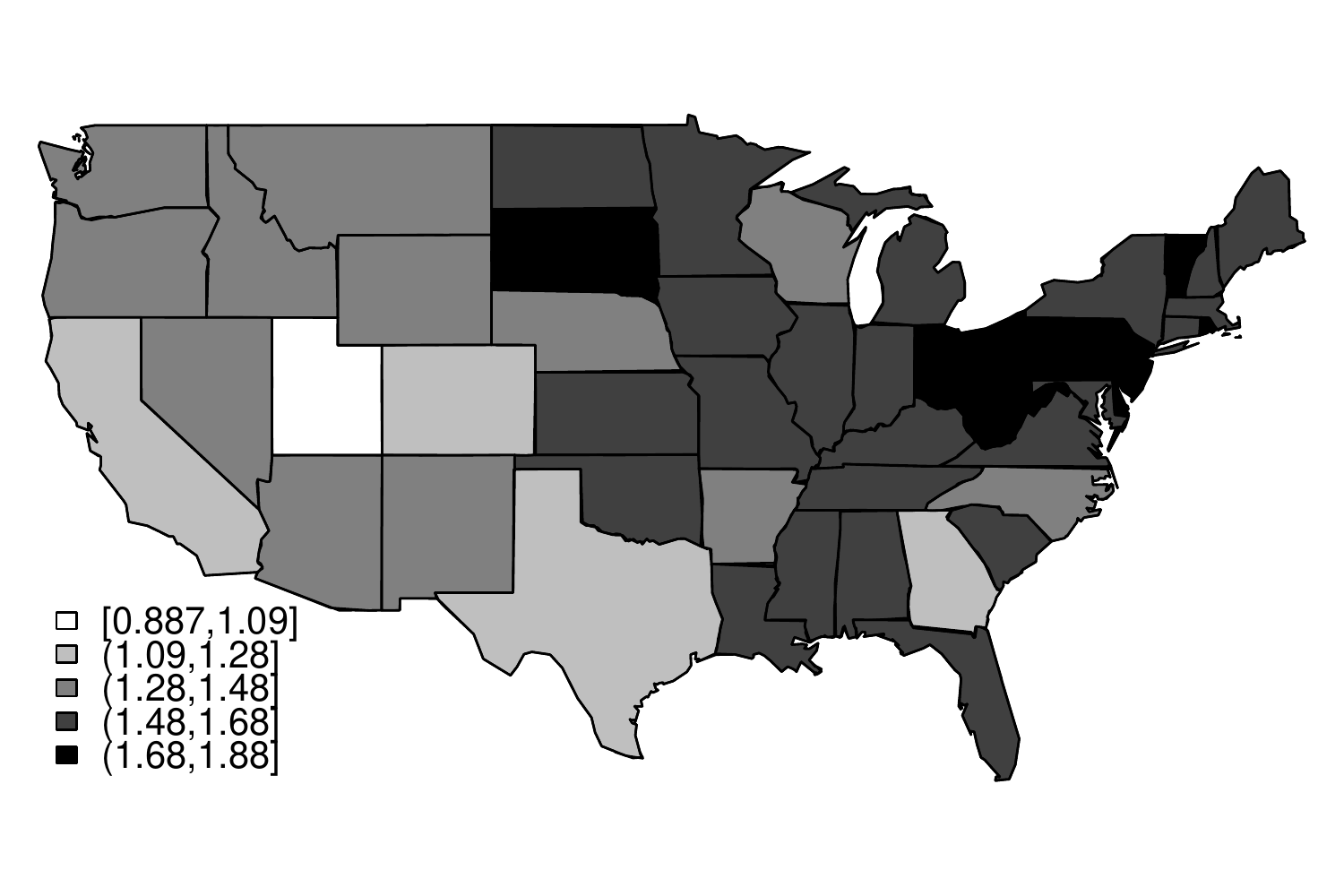}\label{fi:breastraw}}
\subfigure[Prostate]{\includegraphics[width=3.1in,angle=0]{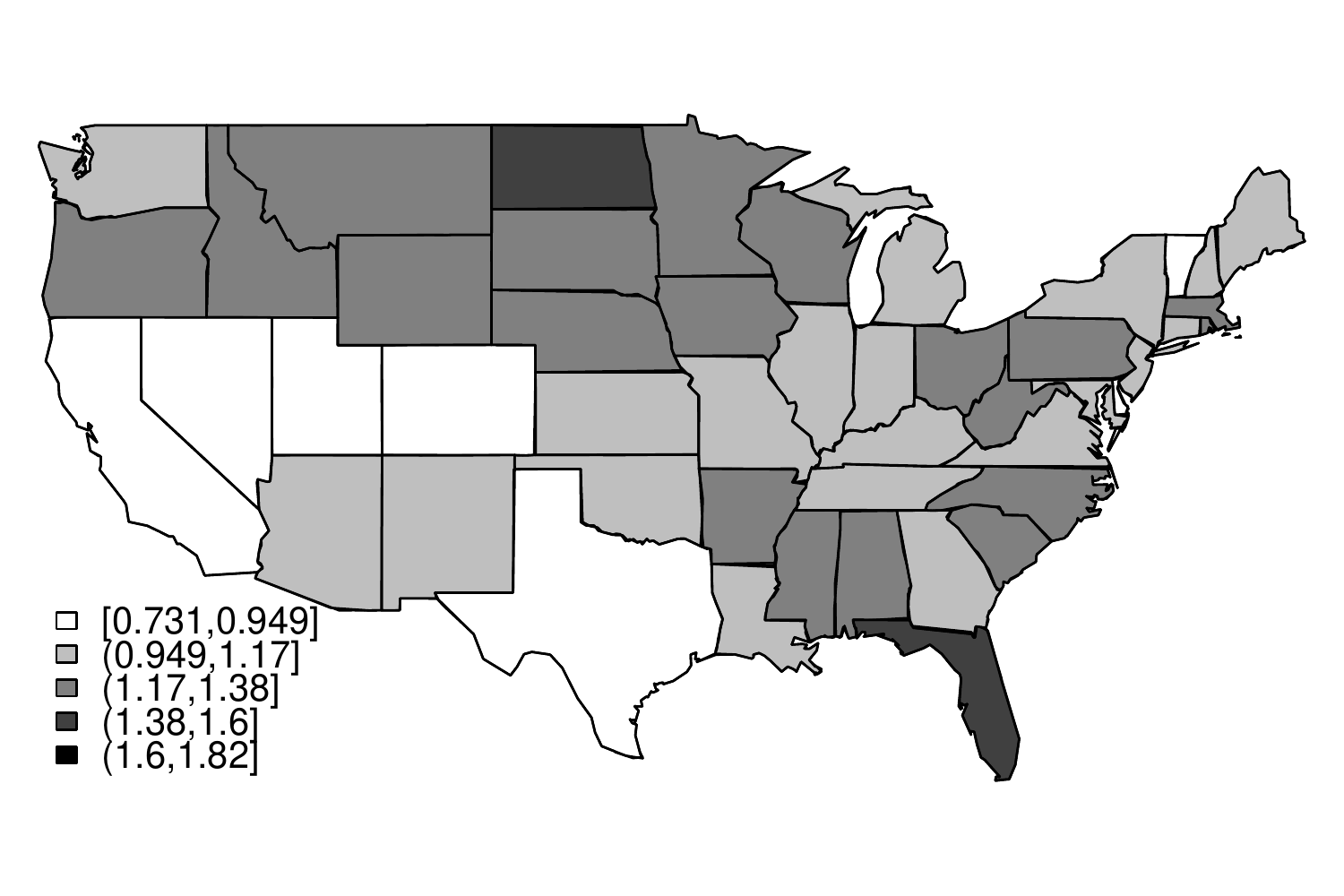}\label{fi:prostateraw}}
\caption{Mortality rates (per 10,000 habitants) in the 48 contiguous states corresponding to four common cancers during 2000.}\label{fi:cancerawmaps}
\end{center}
\end{figure}

The data we analyze consists of mortality counts for 11 types of tumors recorded on the 50 states plus the District of Columbia.  Along with mortality counts, we obtained population counts in order to model death risk.  Figure \ref{fi:cancerawmaps} shows raw mortality rates for four of the most common types of tumors (colon, lung, breast and prostate).  Although the pattern is slightly different for each of these cancers, a number of striking similarities are present; for example, Colorado appears to be a state with low mortality for all of these common cancers, while West Virginia, Pennsylvania and Florida present relatively high mortality rates across the board.

\begin{figure}\caption{Graphical model from the cancer example.  The eleven cancers are: brain (BN), breast (BR), colon (CO), kidney (KD), larynx (LA), lung (LU), oral (OR), ovarian (OV), prostrate (PR), skin (SK), and rectum (RE)}\label{fi:cancergraph}
\begin{center}\includegraphics[width=3.6in, angle=0]{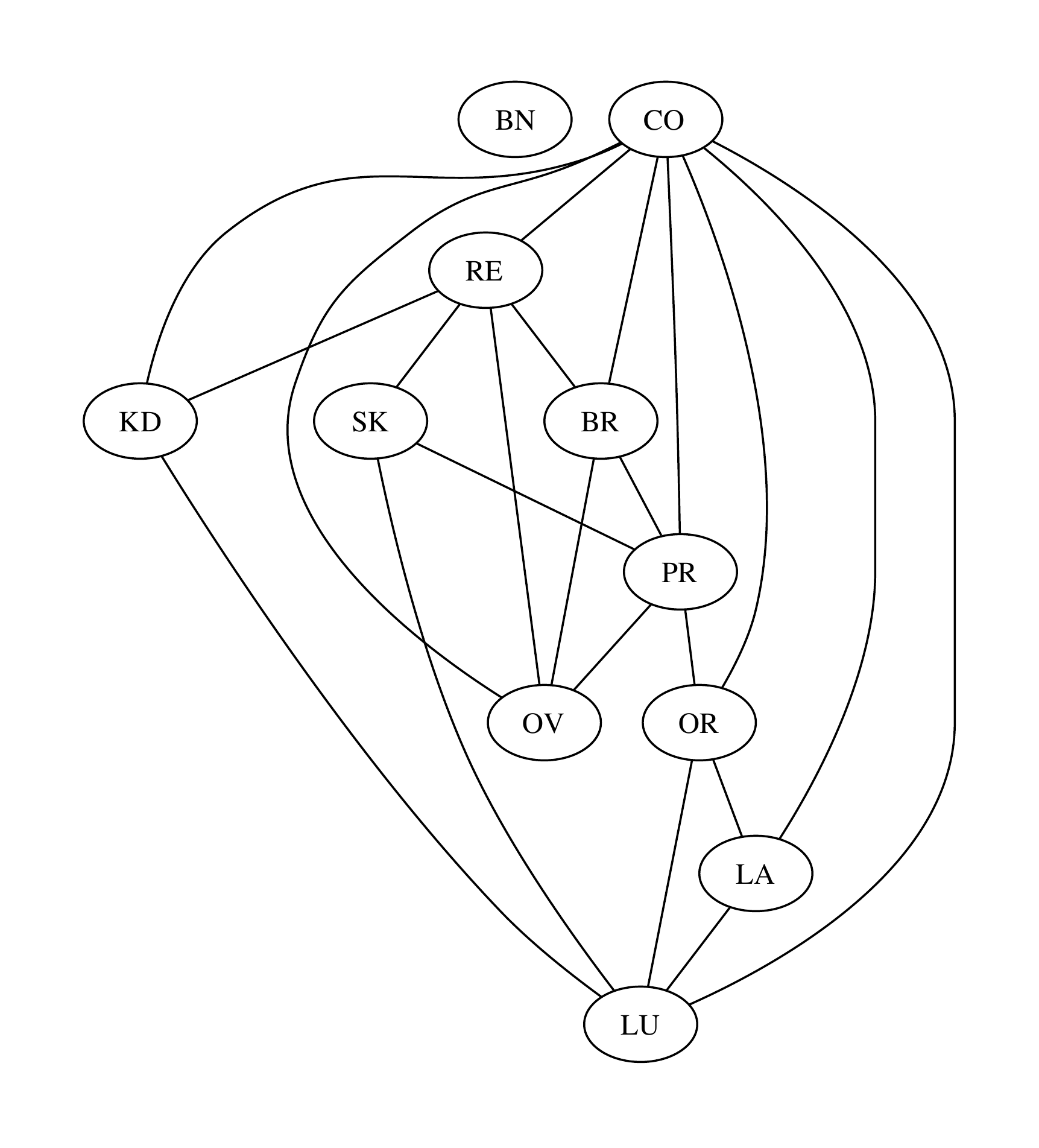}\end{center}
\end{figure}

\begin{figure}[!h]
\begin{center}
\subfigure[Colon]{\includegraphics[width=3.1in,angle=0]{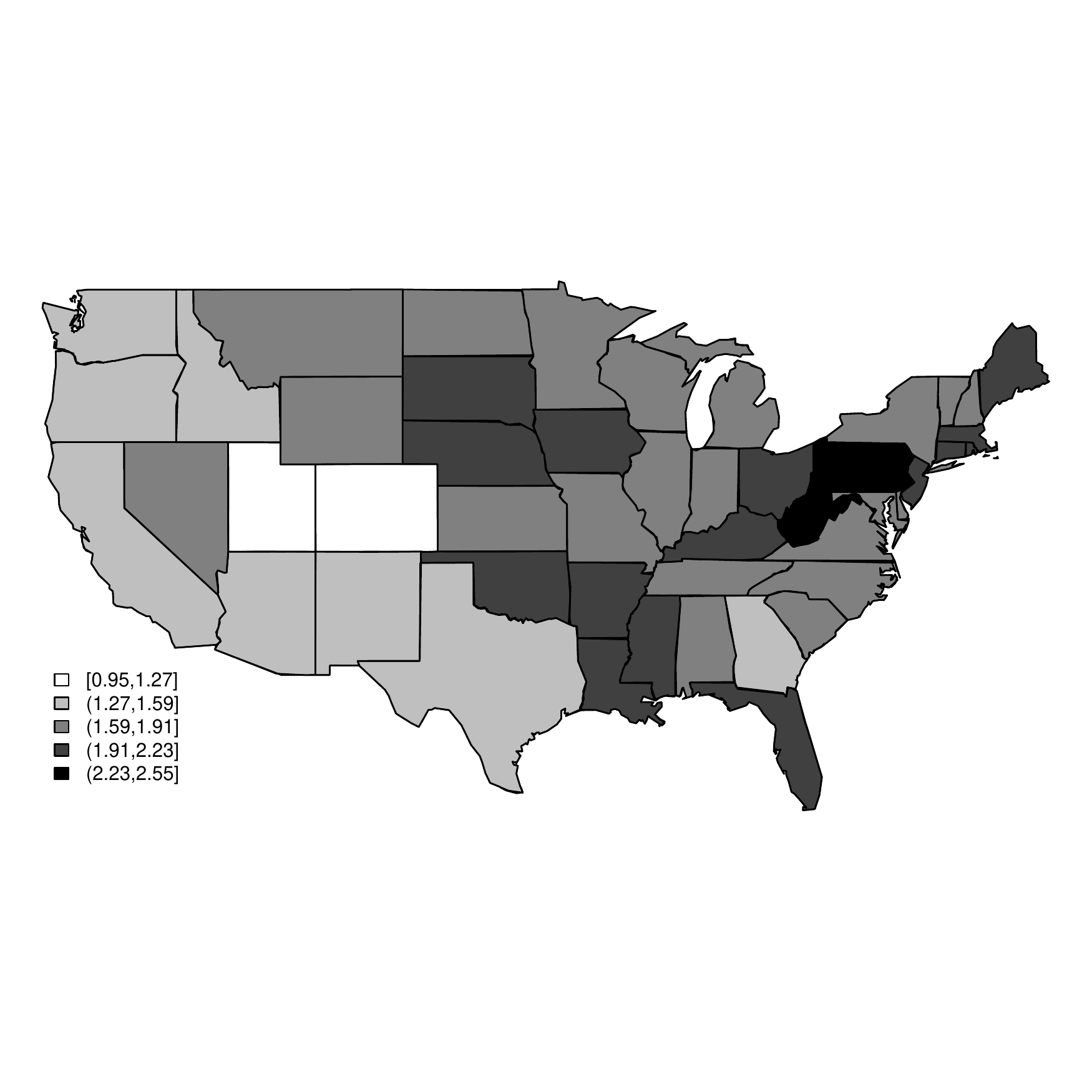}\label{fi:colonfitted}} 
\subfigure[Lung]{\includegraphics[width=3.1in,angle=0]{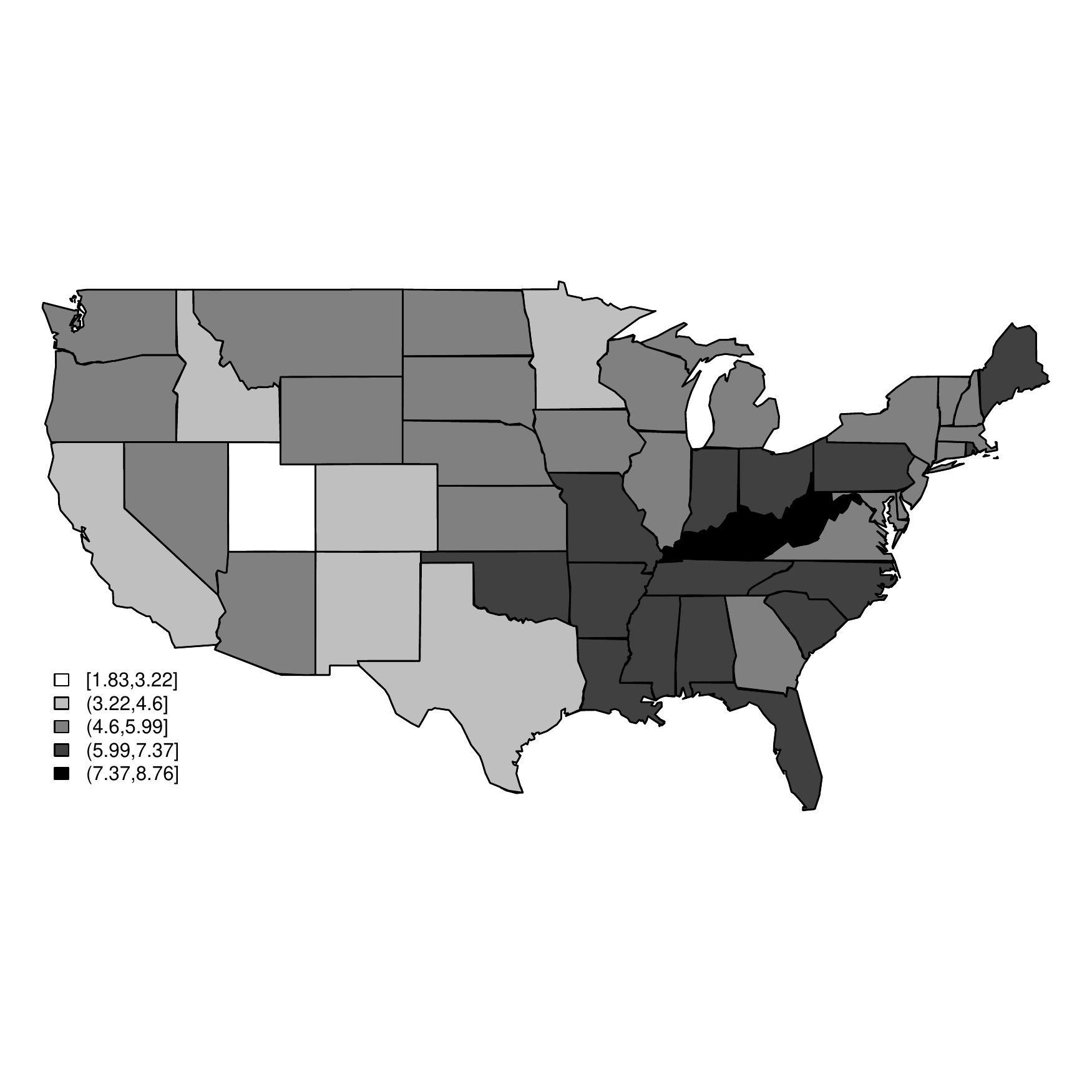}\label{fi:lungfitted}}
\subfigure[Breast]{\includegraphics[width=3.1in,angle=0]{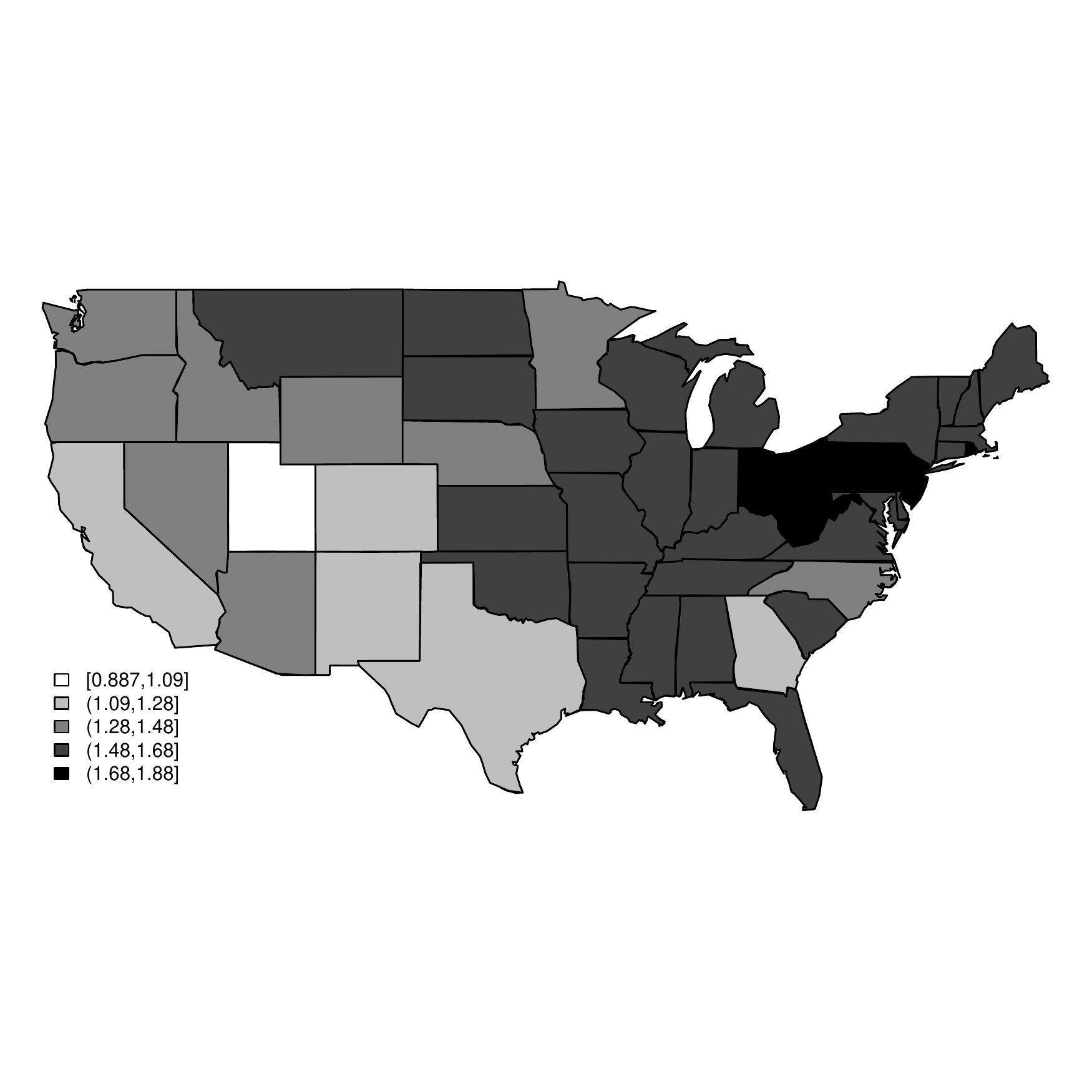}\label{fi:breastfitted}}
\subfigure[Prostate]{\includegraphics[width=3.1in,angle=0]{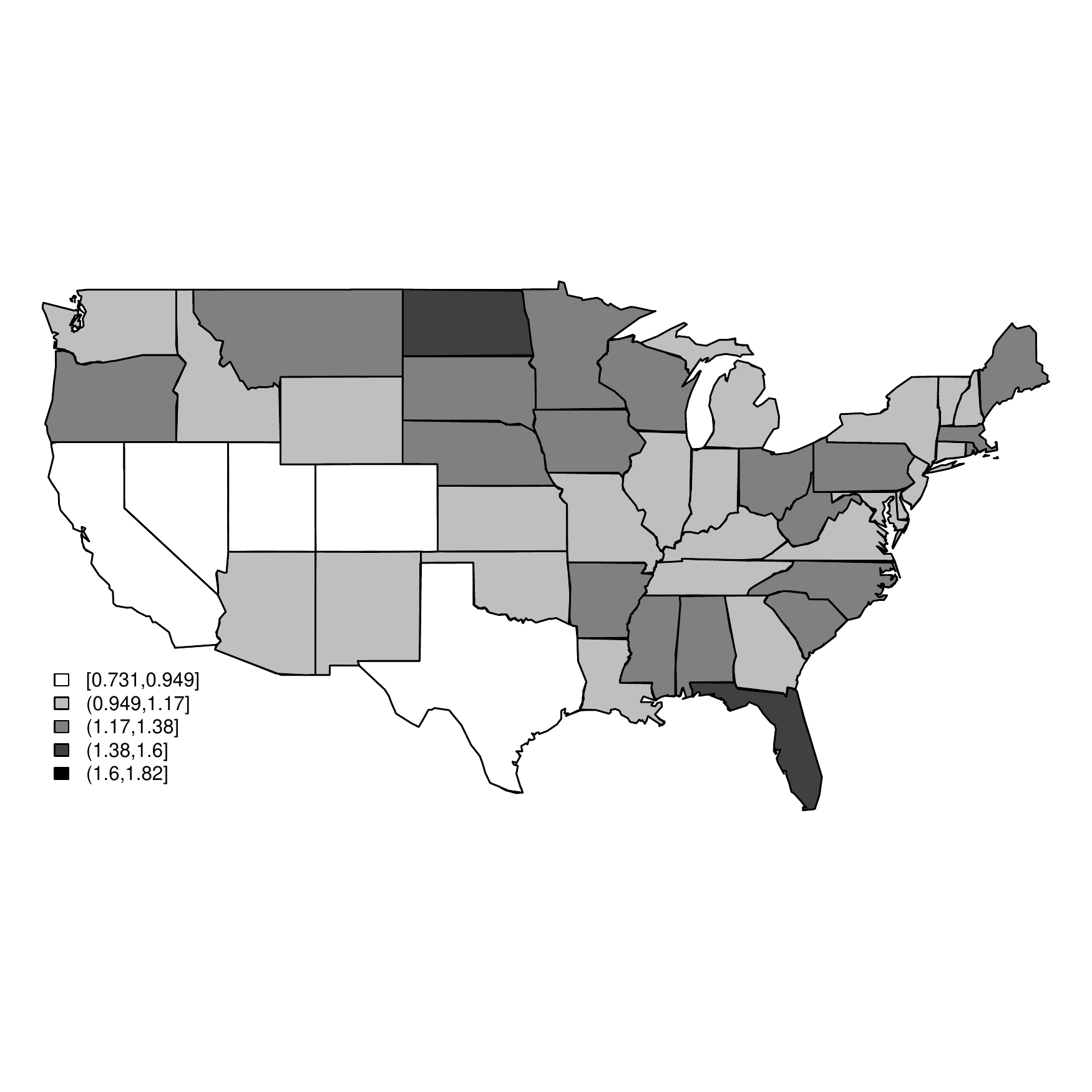}\label{fi:prostatefitted}}
\caption{Fitted Mortality rates (per 10,000 habitants) in the 48 contiguous states corresponding to four common cancers during 2000.}\label{fi:fittedcancers}
\end{center}
\end{figure}

\begin{figure}\caption{Convergence plots for the SAT (left) and cancer examples (right).  These plots show the mean of $z$, by log iteration for each of ten chains.}\label{fi:converge}
\includegraphics[width=3.1in, angle=0]{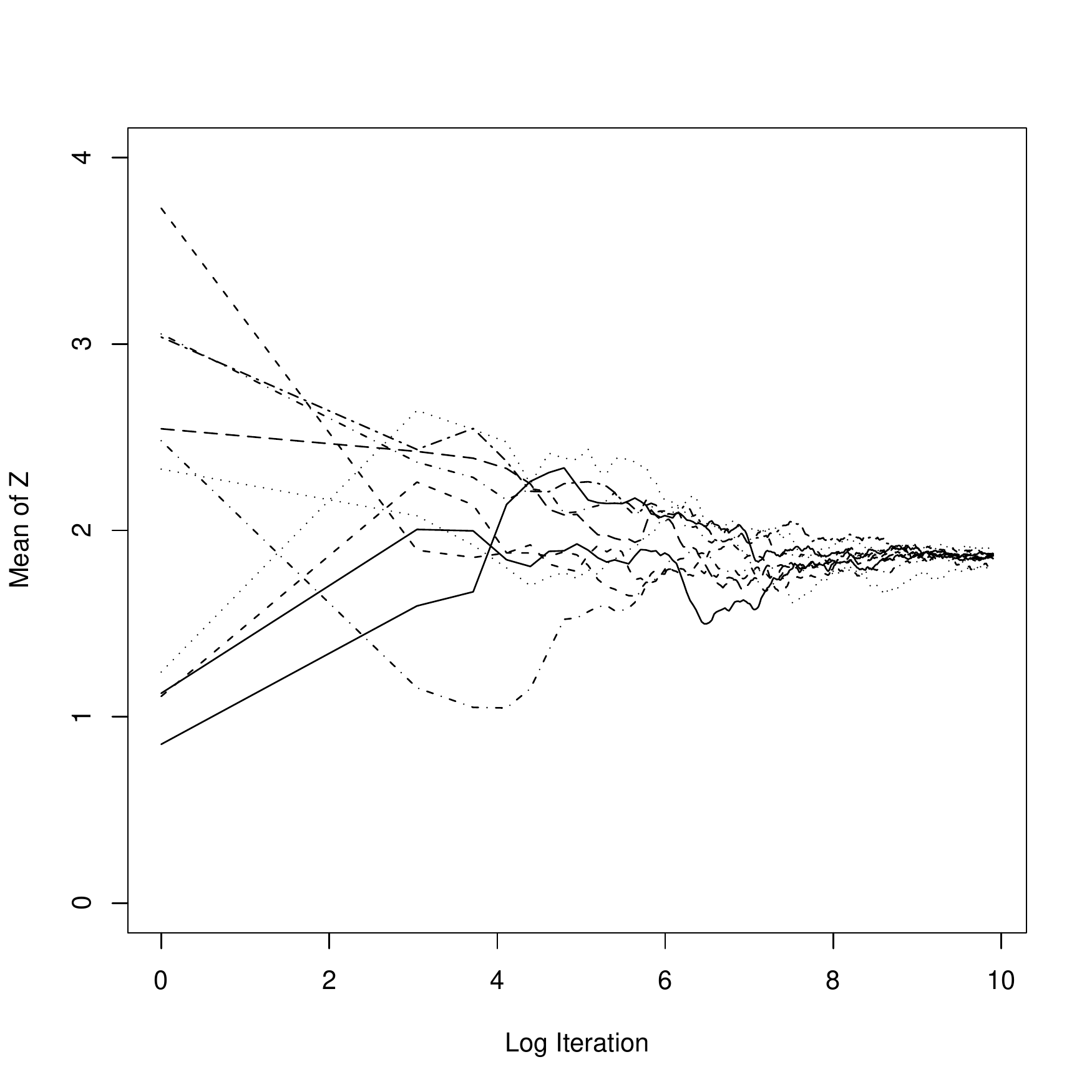}
\includegraphics[width=3.1in, angle=0]{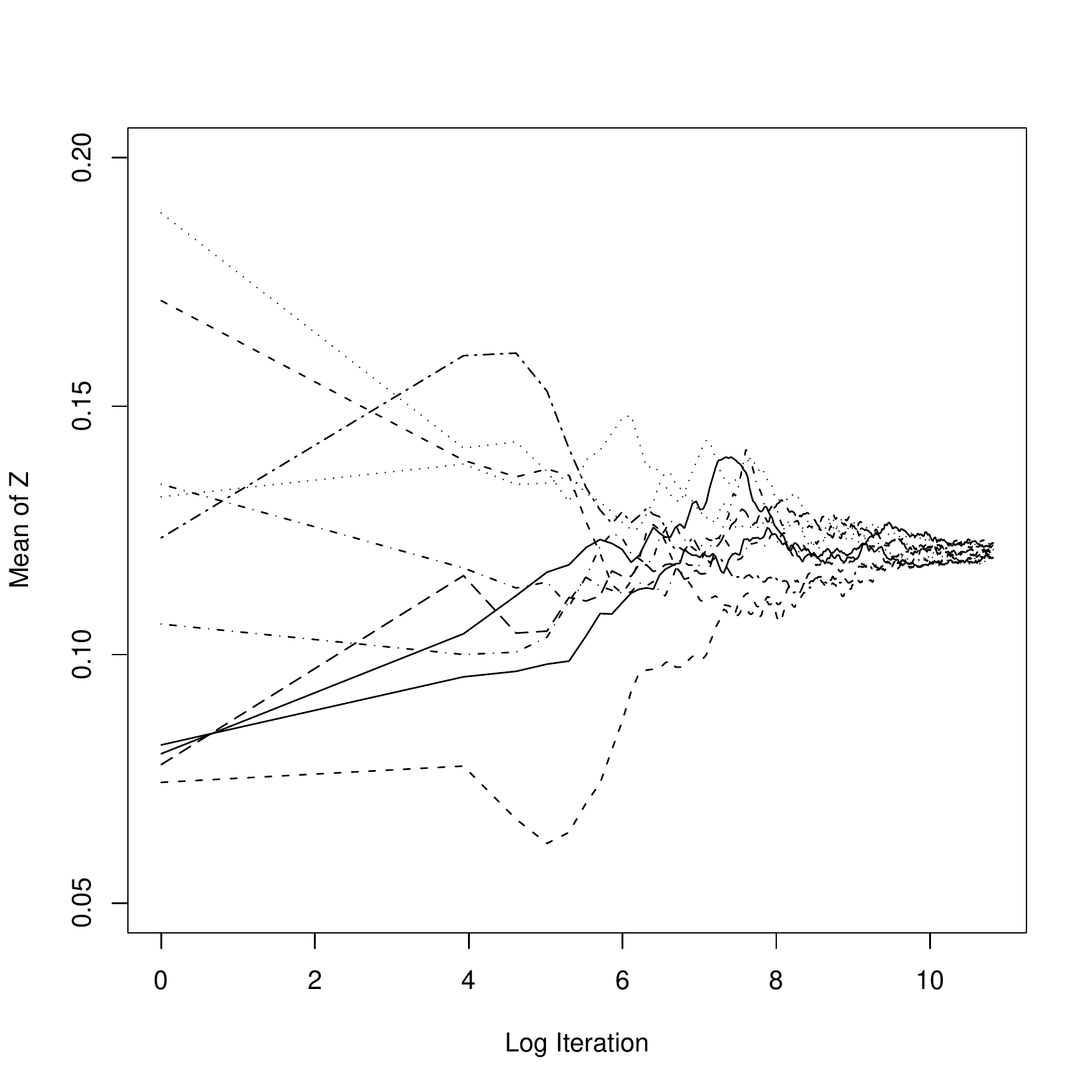}
\end{figure}

A sparse Poisson multivariate loglinear model with spatial random effects was fitted to this data.  The model is constructed along the lines described in Section \ref{sec:spatialglm}.  In particular, we let $Y_{ij}$ be the number of deaths in state $i=1,\ldots,p_{R}=51$ for tumor type $j=1,\ldots,p_{C}=25$. We set 
\begin{eqnarray} \label{eq:datadist}
 Y_{ij} | \eta_{ij} & \sim & \Poi\left( \eta_{ij}\right),\\
 \log\left( \eta_{ij}\right) & = & \log\left( m_{i}\right) + \mu_{j} + X_{ij},\nonumber
\end{eqnarray}
\noindent Here $m_i$ is the population of state $i$, $\mu_j$ is the intercept for tumor type $j$, and $X_{ij}$ is a zero-mean spatial random effect associated with location $i$ and tumor $j$.  We follow the notation from Section \ref{sec:matrixggm} and denote $\widetilde{X}_{ij}=\mu_{j}+X_{ij}$. We further model $\widetilde{\bfX}=\left( \widetilde{X}_{ij}\right)$ with a matrix-variate Gaussian graphical model prior:
\begin{eqnarray}\label{eq:spatialpriorpoi}
 \mbox{vec}\left( \widetilde{\bfX}^{T}\right)|\bfmu,\bfK_{R},\bfK_{C} & \sim & \normal_{p_{R}p_{C}}\left( \mbox{vec}\left\{ \left(\mathbf{1}_{p_{R}} \bfmu^{T}\right)^{T}\right\} ,\left[\bfK_{R}\otimes \bfK_{C}\right]^{-1}\right),\\
 \mathbf{K}_{R}|\delta_{R},\bfD_{R} & \sim & \GWis_{G_{R}}\left(\delta_{R},\bfD_{R}\right),\nonumber\\
 \left(z\mathbf{K}_{C}\right)|\delta_{C},\bfD_{C} & \sim & \GWis_{G_{C}}\left(\delta_{C},\bfD_{C}\right).\nonumber
\end{eqnarray}
Here $\mathbf{1}_{l}$ is the column vector of length $l$ with all elements equal to $1$, $\bfmu = (\mu_1,\ldots,\mu_{p_C})^{T}$ and $z>0$  is an auxiliary variable needed to impose the identifiability constraint $\left( \bfK_{C}\right)_{11}=1$. The row graph $G_R$ is again fixed and derived from the incidence matrix $\bfW$ derived from neighborhood structure across U.S. states, while the column graph $G_{C}$ is given a uniform distribution.  The degrees of freedom for the G-Wishart priors are set as $\delta_{R} = \delta_{C} = 3$, while the centering matrices are chosen as $\bfD_C = \mathbf{I}_{p_C}$ and $\bfD_R = (\delta_{R}-2)w_{1+} ( \bfE_{\bfW} - 0.99\bfW)^{-1}$. We complete the model specification by choosing a uniform prior for the column graph $p\left(G_{C}\right) \propto 1$ and a multivariate normal prior for the mean rates vector $\bfmu \sim\normal_{p_{C}}\left(\bfmu_{0}, \bfOmega^{-1}\right)$ where $\bfmu_{0}=\mu_0 \mathbf{1}_{p_C}$ and $\bfOmega = \omega^{-2} \mathbf{I}_{p_C}$. We set $\mu_{0}$ to the median log incidence rate across all cancers and regions, and $\omega$ to be twice the interquartile range in raw log incidence rates.\\
\indent Posterior inferences on this model are obtained by extending the sampling algorithm described in Section \ref{sec:matrixggm}; details are presented in Appendix \ref{ap:MCMCpoissonglm}.  Since mortality counts below 25 are often deemed unreliable in the cancer surveillance community, we treated them as missing and sampled them as part of our algorithm.  Inferences are based on 50000 samples obtained after a discarding the first 5000 samples.  We ran ten separate chains to monitor convergence.  Figure~\ref{fi:converge} shows the mean value of $z$ by log iteration for each of the ten chains.  All ten agree by 50000 iterations.\\
\indent Figure~\ref{fi:cancergraph} shows the graphical model constructed by adding an edge between any two variables if this edge was present in more than $50\%$ of the total repetitions.  This figure shows several interesting patterns.  First, we see that lung cancer (LU) and colon cancer (CO) have the highest degree of connectivity.  This is a useful result, since lung and colon cancer are the two most prevalent cancers in our dataset.  This implies that variability in these cancers, after controlling for spatial effects, is a good predictor of variability in other lower count cancers.  We also see a clique formed between lung, larynx (LA) and oral (OR) cancers, which is not surprising, as many cases of these cancers are believed to be related to chronic smoking.  Another connexion we expected to see corresponds to the edge between breast (BR) and ovarian (OV) cancer.  Indeed, both of these cancers have been related to mutations in the BRAC genes, and it therefore is reasonable that increased levels of one would be correlated to an increase in the other.  We find that Brain (BN) cancer rates are not correlated with the rates of any other cancer.  Finally, Figure~\ref{fi:fittedcancers} shows the average fitted values for the cancer mortality by state for the four cancers shown in Figure~\ref{fi:cancerawmaps}.  In a result similar to those show in the SAT example above, we see some shrinkage towards the mean, but a preservation of the spatial variability present in the underlying data.\\
\indent Table~\ref{tab:impute}  shows the distribution of the imputed counts for those cancer/state combinations where the reported count was below $25$.

\begin{table}\caption{Distribution of imputed counts for cancers with fewer than $25$ reported counts in a given state}\label{tab:impute}
\begin{center}\begin{tabular}{lccc}
\hline\hline
State & Cancer & Median & 95\% Interval\\
\hline
District of Columbia & Brain & 13 & (3 , 24)\\
North Dakota & Brain & 19 & (9 , 25)\\
District of Columbia & Rectum & 19 & (11 , 24)\\
North Dakota & Rectum & 20 & (13 , 25)\\
Wyoming & Rectum & 15 & (9 , 22)\\
Delaware & Skin & 20 & (12 , 25)\\
District of Columbia & Skin & 15 & (7 , 23)\\
North Dakota & Skin & 17 & (10 , 23)\\
South Dakota & Skin & 20 & (13 , 25)\\
Vermont & Skin & 17 & (10 , 24)\\
Wyoming & Skin & 13 & (8 , 20)\\
District of Columbia & Kidney & 21 & (14 , 25)\\
Wyoming & Kidney & 20 & (13 , 25)\\
Delaware & Oral & 19 & (12 , 25)\\
North Dakota & Oral & 15 & (9 , 22)\\
Rhode Island & Oral & 22 & (16 , 25)\\
South Dakota & Oral & 18 & (11 , 24)\\
Vermont & Oral & 16 & (10 , 23)\\
Wyoming & Oral & 11 & (6 , 18)\\
Delaware & Larynx & 12 & (6 , 19)\\
District of Columbia & Larynx & 11 & (5 , 20)\\
Idaho & Larynx & 12 & (6 , 18)\\
Maine & Larynx & 20 & (11 , 25)\\
Montana & Larynx & 10 & (4 , 16)\\
Nebraska & Larynx & 18 & (11 , 25)\\
New Hampshire & Larynx & 17 & (11 , 24)\\
New Mexico & Larynx & 13 & (7 , 21)\\
North dakota & Larynx & 6 & (2 , 11)\\
Rhode Island & Larynx & 16 & (9 , 23)\\
South Dakota & Larynx & 8 & (4 , 14)\\
Utah & Larynx & 8 & (3 , 15)\\
Vermont & Larynx & 9 & (4 , 15)\\
Wyoming & Larynx & 5 & (2 , 9)\\
\hline\hline
\end{tabular}\end{center}
\end{table}

\section{Discussion}\label{sec:conclusions}

This paper has developed and illustrated a surprisingly powerful computational approach for multivariate and matrix-variate GGMs, and we believe that our application to the construction of spatial models for lattice data makes for particularly appealing illustrations. On one hand, the examples we have considered suggest that the additional flexibility in the spatial correlation structure provided by out approach is necessary to accurately model some spatial data sets.  Indeed, our approach allows for ``non-stationary'' CAR models, where the spatial autocorrelation and spatial smoothing parameters vary spatially.  On the other hand, to be best of our knowledge, we are unaware of any approach in the literature to construct and estimate sparse MCAR, particularly under a Bayesian approach.  In addition to providing insights into the mechanisms underlying the data generation process, it has been repeatedly argued in the literature (see, for example, \citealp{CaWe07} and \citealp{RoDoLe10}) that sparse models tend to provide improved predictions, which is particularly important in such a levant application as cancer mortality surveillance.

The proposed algorithms work well in the examples discussed here where the number of geographical areas is moderate (that is, around fifty or less).  Convergence seems to be achieved quickly (usually within the first 5000 iterations) and running times, although longer than for conventional MCAR models, are still short enough as to make routine implementation feasible (about 1 hour for the simpler SAT dataset, and about 1.5 days for the more complex cancer dataset, both running on a dual-core 2.8 Ghz computer running Linux).  However, computation can still be very challenging when the number of units in the lattice is very high.  In the future we plan to explore parallel implementations (both on regular clusters and GPU cards), and to exploit the sparse structure of the variance-covariance matrix to reduce computational times.

\appendix

\section{Details of the MCMC algorithm for the sparse multivariate spatial Gaussian model for SAT scores}\label{ap:MCMCsat}

The Markov chain Monte Carlo sampler for the spatial regression model in Section \ref{se:sat} involves iterative updating of model parameters from six full conditional distributions.\\
$ $\\
{\bf Step 1: Resample the regression coefficients  $\bfbeta_1 = (\beta_{01}, \beta_{11}, \beta_{21})$}. We denote by $\bfY_{\textasteriskcentered j}$ the $j$-th column of the $p_{R}\times p_{C}$ matrix $\bfY$. From Theorem 2.3.11 of \citet{gupta_nagar_2000} applied to equation (\ref{eq:ysat}) we have
$$
\bfY_{\textasteriskcentered 1}|\bfY_{\textasteriskcentered 2} \sim \normal_{p_{R}}\left( \bfZ \bfbeta_{1}+\left( \bfY_{\textasteriskcentered 2} -  \bfZ \bfbeta_{2}\right)\frac{\left( \bfK_{C}^{-1}\right)_{21}}{\left( \bfK_{C}^{-1}\right)_{22}},\left[ \left(\bfK_{C}\right)_{11}\bfK_{R}\right]^{-1}\right).
$$
Given the prior $\bfbeta_{2}\sim \normal_{3}\left( \bfb_{2},\bfOmega_{2}^{-1}\right)$ and the constraint $\left( \bfK_{C}\right)_{11} = 1$, it follows that $\bfbeta_{1}$ is updated by sampling from the multivariate normal $\normal_{3}\left( \mathbf{m}_{\bfbeta_{2}},\bfOmega_{\bfbeta_{2}}^{-1}\right)$ where
\begin{eqnarray*}
 \bfOmega_{\bfbeta_{2}} = \bfZ^{T}\bfK_{R}\bfZ + \bfOmega_{2},\quad \mathbf{m}_{\bfbeta_{2}} = \bfOmega_{\bfbeta_{2}}^{-1}\left\{  \bfZ^{T}\bfK_{R}\left[ \bfY_{\textasteriskcentered 1} - \left( \bfY_{\textasteriskcentered 2} -  \bfZ \bfbeta_{2}\right)\frac{\left( \bfK_{C}^{-1}\right)_{21}}{\left( \bfK_{C}^{-1}\right)_{22}}\right] + \bfOmega_{2}\bfb_{2}\right\}.
\end{eqnarray*}
{\bf Step 2: Resample the regression coefficients  $\bfbeta_2 = (\beta_{02}, \beta_{12}, \beta_{22})$}. Since apriori $\bfbeta_{1}\sim \normal_{3}\left( \bfb_{1},\bfOmega_{1}^{-1}\right)$, it follows that $\bfbeta_{1}$ is updated by sampling from the multivariate normal $\normal_{3}\left( \mathbf{m}_{\bfbeta_{1}},\bfOmega_{\bfbeta_{1}}^{-1}\right)$ where
\begin{eqnarray*}
 \bfOmega_{\bfbeta_{1}} = \left( \bfK_{C}\right)_{22} \bfZ^{T}\bfK_{R}\bfZ + \bfOmega_{1},\quad
 \mathbf{m}_{\bfbeta_{1}} = \bfOmega_{\bfbeta_{1}}^{-1}\left\{  \left( \bfK_{C}\right)_{22} \bfZ^{T}\bfK_{R}\left[ \bfY_{\textasteriskcentered 2} - \left( \bfY_{\textasteriskcentered 1} -  \bfZ \bfbeta_{1}\right)\frac{\left( \bfK_{C}^{-1}\right)_{12}}{\left( \bfK_{C}^{-1}\right)_{11}}\right] + \bfOmega_{1}\bfb_{1}\right\}.
\end{eqnarray*}
{\bf Step 3: Resample the row precision matrix $\bfK_R$ encoding the spatial structure}.\\
{\bf Step 4: Resample the column graph $G_C$}.\\
{\bf Step 5: Resample the column precision matrix $\bfK_C$}.\\
{\bf Step 6:  Resample the auxiliary variable $z$}.\\
After updating the regression coefficients in Steps 1 and 2, we recalculate the residual matrix $\bfX = \left( X_{ij}\right)$, where $X_{ij} = Y_{ij} - \beta_{0j} - \beta_{1j}Z_i - \beta_{2j}Z_{i}^2$. The updated $\bfX$ represents a matrix variate dataset with a sample size $n=1$, based on which $\bfK_R$, $G_C$, $\bfK_C$ and $z$ are updated. Thus Steps 3, 4, 5 and 6 above correspond to Steps 2, 3, 4 and 5 from Section \ref{sec:matrixggm}, and no new algorithm needs to be described.  Note that Step 1 of Section \ref{sec:matrixggm} is unnecessary as the row graph $G_R$ is assumed known, and given by the neighborhood structure.

\section{Details of the MCMC algorithm for the sparse multivariate spatial Poisson count model}\label{ap:MCMCpoissonglm}

The Markov chain Monte Carlo sampler for the sparse multivariate spatial Poisson count model from Section \ref{se:diseasemap} involves iterative updating of model parameters in six steps.\\
$ $\\
{\bf Step 1: Resample the mean rates $\bfmu$}. The full conditional for $\bfmu$ is
\begin{eqnarray*}
 p\left( \bfmu | \widetilde{\bfX},\bfK_{R},\bfK_{C}\right) \propto \exp\left\{ -\frac{1}{2} \mbox{tr}\left[ \bfK_{R}\left( \widetilde{\bfX} - \mathbf{1}_{p_R}\bfmu^{T} \right) \bfK_{C}\left( \widetilde{\bfX} - \mathbf{1}_{p_R}\bfmu^{T} \right)^{T} +\mathbf{1}_{p_R}\left( \bfmu-\bfmu_{0}\right)^{T} \bfOmega\left( \bfmu-\bfmu_{0}\right)  \mathbf{1}_{p_R}^{T}\right]\right\}.
\end{eqnarray*}
It follows that $\bfmu$ is updated by direct sampling from the multivariate normal $\normal_{p_{C}} \left( \mathbf{m}_{\bfmu},\bfK_{\bfmu}^{-1}\right)$ where
\begin{eqnarray*}
 \bfK_{\bfmu} = \left(\mathbf{1}_{p_R}^{T} \bfK_R \mathbf{1}_{p_R}\right) \bfK_C + p_{R}\bfOmega,\quad
 \mathbf{m}_{\bfmu} = \bfK_{\bfmu}^{-1}\left[ \bfK_C \widetilde{\bfX}^{T} \bfK_R\mathbf{1}_{p_R} + p_{R}\bfOmega\bfmu_0 \right].
\end{eqnarray*}
{\bf Step 2: Resample $\widetilde{\bfX}$}. We update the centered random effects $\widetilde{\bfX}$ by sequentially updating each row vector $\widetilde{\bfX}_{i\textasteriskcentered}$, $i=1,\ldots,p_{R}$. The matrix-variate GGM prior from equation (\ref{eq:spatialpriorpoi}) implies the following conditional prior for $\widetilde{\bfX}_{i\textasteriskcentered}$ \citep{gupta_nagar_2000}:
\begin{eqnarray}\label{eq:proposal1}
 \widetilde{\bfX}_{i\textasteriskcentered}^{T}| \left\{ \widetilde{\bfX}_{i^{\prime}\textasteriskcentered}:i^{\prime}\ne i \right\},\bfmu,\bfK_{R},\bfK_{C} & \sim & \normal_{p_{C}} \left( \mathbf{M}_{i},\bfOmega_{i}^{-1}\right),
\end{eqnarray}
where
$$
 \mathbf{M}_{i} = \bfmu - \sum_{i^{\prime} \in \partial i} \frac{(\bfK_R)_{ii^{\prime}}}{(\bfK_R)_{ii}} \left(\widetilde{\bfX}_{i^{\prime}\textasteriskcentered}^{T} -\bfmu \right),\quad \bfOmega_{i} = \left( \bfK_{R}\right)_{ii}\bfK_{C}.
$$
It follows that the full conditional distribution of $\widetilde{\bfX}_{i\textasteriskcentered}$ is written as
\begin{multline}\label{eq:fullcond_step1}
p\left(\widetilde{\bfX}_{i\textasteriskcentered} | \left\{ \widetilde{\bfX}_{i^{\prime}\textasteriskcentered} : i' \ne i  \right\}, \bfY_{i\textasteriskcentered},\bfmu,\bfK_{R},\bfK_{C}\right) \propto \\
\propto \left[ \prod_{j=1}^{p_{C}}  \exp\left\{  Y_{ij}\left( \log(m_{i}) + \widetilde{X}_{ij} \right) - m_{i} \exp\left( \widetilde{X}_{ij} \right)  \right\}  \right] \exp\left\{  -\frac{1}{2} \left[ \widetilde{\bfX}_{i\textasteriskcentered} - \left( \mathbf{M}_{i}\right)^{T}\right] \bfOmega_{i}  \left[ \left(\widetilde{\bfX}_{i\textasteriskcentered}\right)^{T} - \mathbf{M}_{i}\right]
\right\}.
\end{multline}
Since the distribution (\ref{eq:fullcond_step1}) does not correspond to any standard distribution, we design a Metropolis-Hastings algorithm to sample from it.  We consider a strictly positive precision parameter $\widetilde{\sigma}$. For each $j=1,\ldots,p_{C}$, we update the $j$-th element of $\widetilde{\bfX}_{i\textasteriskcentered}$ by sampling $\gamma\sim \normal\left( \widetilde{X}_{ij},\widetilde{\sigma}^{2}\right)$. We define a candidate row vector $\widetilde{\bfX}^{new}_{i\textasteriskcentered}$ by replacing $\widetilde{X}_{ij}$ with $\gamma$ in $\widetilde{\bfX}_{i\textasteriskcentered}$. We update the current $i$-th row of $\widetilde{\bfX}$ with $\widetilde{\bfX}^{new}_{i\textasteriskcentered}$ with probability
$$
 \min \left\{ 1, \frac{p\left(\widetilde{\bfX}_{i\textasteriskcentered}^{new} | \left\{ \widetilde{\bfX}_{i^{\prime}\textasteriskcentered} : i' \ne i  \right\}, \bfY_{i\textasteriskcentered},\bfmu,\bfK_{R},\bfK_{C}\right)}{p\left(\widetilde{\bfX}_{i\textasteriskcentered} | \left\{ \widetilde{\bfX}_{i^{\prime}\textasteriskcentered} : i' \ne i  \right\}, \bfY_{i\textasteriskcentered},\bfmu,\bfK_{R},\bfK_{C}\right)}\right\}.
$$
Otherwise the $i$-th row of $\widetilde{\bfX}$ remains unchanged.\\
{\bf Step 3: Resample the row precision matrix $\bfK_R$ encoding the spatial structure}.\\
{\bf Step 4: Resample the column graph $G_C$}.\\
{\bf Step 5: Resample the column precision matrix $\bfK_C$}.\\
{\bf Step 6:  Resample the auxiliary variable $z$}.\\
After completing Steps 1 and 2, we recalculate the zero-mean spatial random effects $X_{ij} = \widetilde{X}_{ij}-\mu_{j}$. The updated $\bfX = \left( X_{ij}\right)$ represents a matrix variate dataset with a sample size $n=1$, based on which $\bfK_R$, $G_C$, $\bfK_C$ and $z$ are updated. Steps 3, 4, 5 and 6 above are Steps 2, 3, 4 and 5 from Section \ref{sec:matrixggm}.

\bibliographystyle{bka}
\bibliography{car-ggm-03212010}

\end{document}